# GeoJSEval: An Automated Evaluation Framework for Large Language Models on JavaScript-Based Geospatial Computation and Visualization Code Generation

Guanyu Chen[a], Haoyue Jiao[a], Shuyang Hou[b], Ziqi Liu[b]*, Lutong Xie[b], Shaowen Wu[b], Huayi Wu[b], Xuefeng Guan[b], Zhipeng Gui[c]

a. School of Resource and Environmental Sciences, Wuhan University, Wuhan, China

*b. State Key Laboratory of Information Engineering in Surveying Mapping and Remote Sensing, Wuhan University, Wuhan, China

c. School of Remote Sensing and Information Engineering, Wuhan University, Wuhan, China

*Corresponding author: Ziqi Liu, email: lzq677@whu.edu.cn

## Abstract

With the widespread adoption of large language models (LLMs) in code generation tasks, geospatial code generation has emerged as a critical frontier in the integration of artificial intelligence and geoscientific analysis. This growing trend underscores the urgent need for systematic evaluation methodologies to assess the generation capabilities of LLMs in geospatial contexts. In particular, geospatial computation and visualization tasks in the JavaScript environment rely heavily on the orchestration of diverse frontend libraries and ecosystems, posing elevated demands on a model's semantic comprehension and code synthesis capabilities. To address this challenge, we propose GeoJSEval—the first multimodal, function-level automatic evaluation framework for LLMs in JavaScript-based geospatial code generation tasks. The framework comprises three core components: a standardized test suite (GeoJSEval-Bench), a code submission engine, and an evaluation module. It includes 432 function-level tasks and 2,071 structured test cases, spanning five widely used JavaScript geospatial libraries that support spatial analysis and visualization functions, as well as 25 mainstream geospatial data types. GeoJSEval enables multidimensional quantitative evaluation across metrics such as accuracy, output stability, resource consumption, execution efficiency, and error type distribution. Moreover, it integrates boundary testing mechanisms to enhance robustness and evaluation coverage. We conduct a comprehensive assessment of 18 state-of-the-art LLMs using GeoJSEval, uncovering significant performance disparities and bottlenecks in spatial semantic understanding, code reliability, and function invocation accuracy. GeoJSEval offers a foundational methodology, evaluation resource, and practical toolkit for the standardized assessment and optimization of geospatial code generation models, with strong extensibility and promising applicability in real-world scenarios.

## 1. Introduction

General-purpose code is typically written in formal programming languages such as Python, C++, or Java, and is employed to perform data processing, implement algorithms, and handle multitask operations such as network communication(Brady, 2013). Users translate logical expressions into executable instructions through programming, enabling the processing of diverse data types including text, structured data, and imagery. However, the rapid growth of high-resolution remote sensing imagery and crowdsourced spatiotemporal data has imposed greater demands on computational capacity and methodological customization in geospatial information analysis, particularly for tasks involving indexing, processing, and analysis of heterogeneous geospatial data. Geospatial data types are inherently complex, encompassing remote sensing imagery, vector layers, and point clouds. These data involve various geometric structures (points, lines, polygons, volumes) and semantic components such as attributes, metadata, and temporal information(Li et al., 2016; Sun et al., 2019). Spatial analysis commonly entails dozens of operations, including buffer generation, point clustering, spatial interpolation, spatiotemporal trajectory analysis, isopleth and isochrone generation, linear referencing, geometric merging and clipping, and topological relationship evaluation(Ding & Fotheringham, 1992). Effective geospatial data processing requires integrated consideration of spatial reference system definition and transformation, spatial indexing structures (e.g., R-trees, Quad-trees), validation of geometric and topological correctness, and the modeling and management of spatial databases(Breunig et al., 2020). Moreover, modern geospatial applications demand the integration of functionalities such as basemap services (e.g., OpenStreetMap, Mapbox), vector rendering, user interaction, heatmaps, trajectory animation, and three-dimensional visualization(Evangelidis et al., 2018; Foerster et al., 2012). Consequently, the foundational libraries of conventional general-purpose languages are increasingly insufficient for addressing the complexities of contemporary geospatial analysis.

Although geospatial analysis platforms such as ArcGIS and QGIS have integrated a wide range of built-in functions, they primarily rely on graphical user interfaces (GUIs) for point-and-click operations. While these interfaces are user-friendly, they present limitations in terms of workflow customization, task reproducibility, and methodological dissemination(Graser & Olaya, 2015; Marinoni, 2004). In contrast, code-driven approaches offer greater flexibility and reusability. Users can automate complex tasks and enable cross-context reuse through scripting, thereby facilitating the efficient dissemination and sharing of GIS methodologies. Meanwhile, cloud-based platforms such as Google Earth Engine (GEE) provide JavaScript and Python APIs that support large-scale online analysis of remote sensing and spatiotemporal data(Zhao et al., 2021). However, the underlying computational logic of these platforms depends on pre-packaged functions and remote execution environments, which restrict fine-grained control over processing workflows. As a result, they struggle to meet demands for local deployment, custom operator definition, and data privacy—particularly in scenarios requiring enhanced security or functional extensibility. In recent years, the rise of WebGIS has been driving a paradigm shift in spatial analysis toward browser-based execution. In applications such as urban governance, disaster response, agricultural monitoring, ecological surveillance, and public participation, users increasingly expect an integrated, browser-based environment for data loading, analysis, and visualization—enabling a lightweight, WYSIWYG (what-you-see-is-what-you-get) experience(Liu & Xu, 2001). This trend is propelling the evolution of geospatial computing from traditional desktop and server paradigms toward "computable WebGIS." The browser is increasingly serving as a lightweight yet versatile computational terminal, opening new possibilities for frontend spatial modeling and visualization.

Against this backdrop, JavaScript has emerged as a key language for geospatial analysis and visualization, owing to its native browser execution capability and mature frontend ecosystem. Compared to languages such as Python (e.g., GeoPandas, Shapely, Rasterio), R, or MATLAB, JavaScript supports direct execution in the browser, eliminating the need for environment setup. This makes it particularly well-suited for lightweight analysis workflows with low resource overhead and strong cross-platform compatibility(Dai et al., 2009). For spatial tasks, the open-source community has developed a rich ecosystem of JavaScript-based geospatial tools, which can be broadly categorized into two types. The first category focuses on spatial analysis and computation. For instance, Turf.js offers operations such as buffering, merging, clipping, and distance measurement; JSTS supports topological geometry processing; and Geolib enables precise calculations of distances and angles. The second category centers on map visualization. Libraries such as Leaflet and OpenLayers provide functionalities for layer management, plugin integration, and interactive user interfaces, significantly accelerating the shift of GIS capabilities to the Web(Jia et al., 2009). In this paper, we collectively refer to these tools as "JavaScript-based Geospatial Code," which enables the execution of core geospatial analysis and visualization tasks without requiring backend services. These tools are particularly suitable for small-to-medium-scale data processing, prototype development, and educational demonstrations. While JavaScript may fall short of Python in terms of large-scale computation or deep learning integration, it offers distinct advantages in frontend spatial computing and method sharing due to its excellent platform compatibility, low development barrier, and high reusability(Dietrich et al., 2016).

Authoring JavaScript-based geospatial code requires not only general programming skills but also domain-specific knowledge of spatial data types, core library APIs, spatial computation logic, and visualization techniques—posing a significantly higher barrier to entry than conventional coding tasks(Hou et al., 2016). As GIS increasingly converges with fields such as remote sensing, urban studies, and environmental science, a growing number of non-expert users are eager to engage in geospatial analysis and application development. However, many of them encounter substantial difficulties in modeling and implementation, highlighting the urgent need for automated tools to lower learning and execution costs. In recent years, large language models (LLMs), empowered by massive code corpora and Transformer-based architectures, have achieved remarkable progress in natural language–driven code generation tasks. Models such as GPT-4o, DeepSeek, Claude, and LLaMA are capable of translating natural language prompts directly into executable program code(Deng et al., 2025). Domain-specific models such as DeepSeek Coder, Qwen 2.5-Coder, and Code LLaMA have further enhanced accuracy and robustness, offering feasible solutions for automated code generation in geospatial tasks(Hou, Shen, Zhao, et al., 2025). However, unlike general-purpose coding tasks, geospatial code generation entails elevated demands in terms of semantic understanding, library function invocation, and structural organization—challenges that are particularly pronounced in the JavaScript environment. Compared to backend-oriented languages, the JavaScript ecosystem is characterized by fragmented APIs, inconsistent documentation, and sensitivity to execution environments. Moreover, it often requires the tight integration of map rendering, spatial computation, and interactive logic, thereby intensifying the need for precise semantics and coherent module organization(Fang et al., 2020). Current mainstream LLMs frequently exhibit hallucination issues when generating JavaScript-based geospatial code. These manifest as incorrect function calls, spatial semantic misunderstandings, flawed analytical workflows, misaligned rendering–interaction logic, parameter-type mismatches, and spatial scale misconfigurations (**Figure 1**), often resulting in erroneous or non-executable outputs.

This raises a critical question: how well do current LLMs actually perform in generating JavaScript-based geospatial code? Without systematic evaluation, direct application of these models may result in code with structural flaws, logical inconsistencies, or execution failures—ultimately compromising the reliability of geospatial analyses. Although LLMs have demonstrated strong performance on general code benchmarks such as HumanEval and MBPP(Wan et al., 2024), these benchmarks overlook the inherent complexity of geospatial tasks and the unique characteristics of the JavaScript frontend ecosystem, making them insufficient for assessing real-world usability and stability in spatial contexts. Existing studies have made preliminary attempts to evaluate geospatial code generation, yet significant limitations remain. For instance, GeoCode-Bench and GeoCode-Eval proposed by Wuhan University(Hou, Shen, Zhao, et al., 2025), and the GeoSpatial-Code-LLMs Dataset developed by Wrocław University of Science and Technology(Quarati et al., 2021), primarily focus on Python or GEE-based code, with limited attention to JavaScript-centric scenarios. Specifically, GeoCode-Bench and GeoCode-Eval rely on manual scoring, resulting in high evaluation costs, subjectivity, and limited reproducibility. While the GeoSpatial-Code-LLMs Dataset introduces automation to some extent, it suffers from a small sample size (approximately 40 examples) and lacks coverage of multimodal data structures such as vectors and rasters. Moreover, AutoGEEval and AutoGEEval++ have established evaluation pipelines for the GEE platform, yet they fail to address browser-based execution environments or the distinctive requirements of JavaScript frontend tasks(Hou, Shen, Wu, et al., 2025; Wu et al., 2025). Therefore, there is an urgent need to develop an automated, end-to-end evaluation pipeline for systematically assessing the performance of mainstream LLMs in JavaScript-based geospatial code generation. Such a framework is essential for identifying model limitations and appropriate application scopes, guiding future optimization and fine-tuning efforts, and laying the theoretical and technical foundation for building low-barrier, high-efficiency geospatial code generation tools.

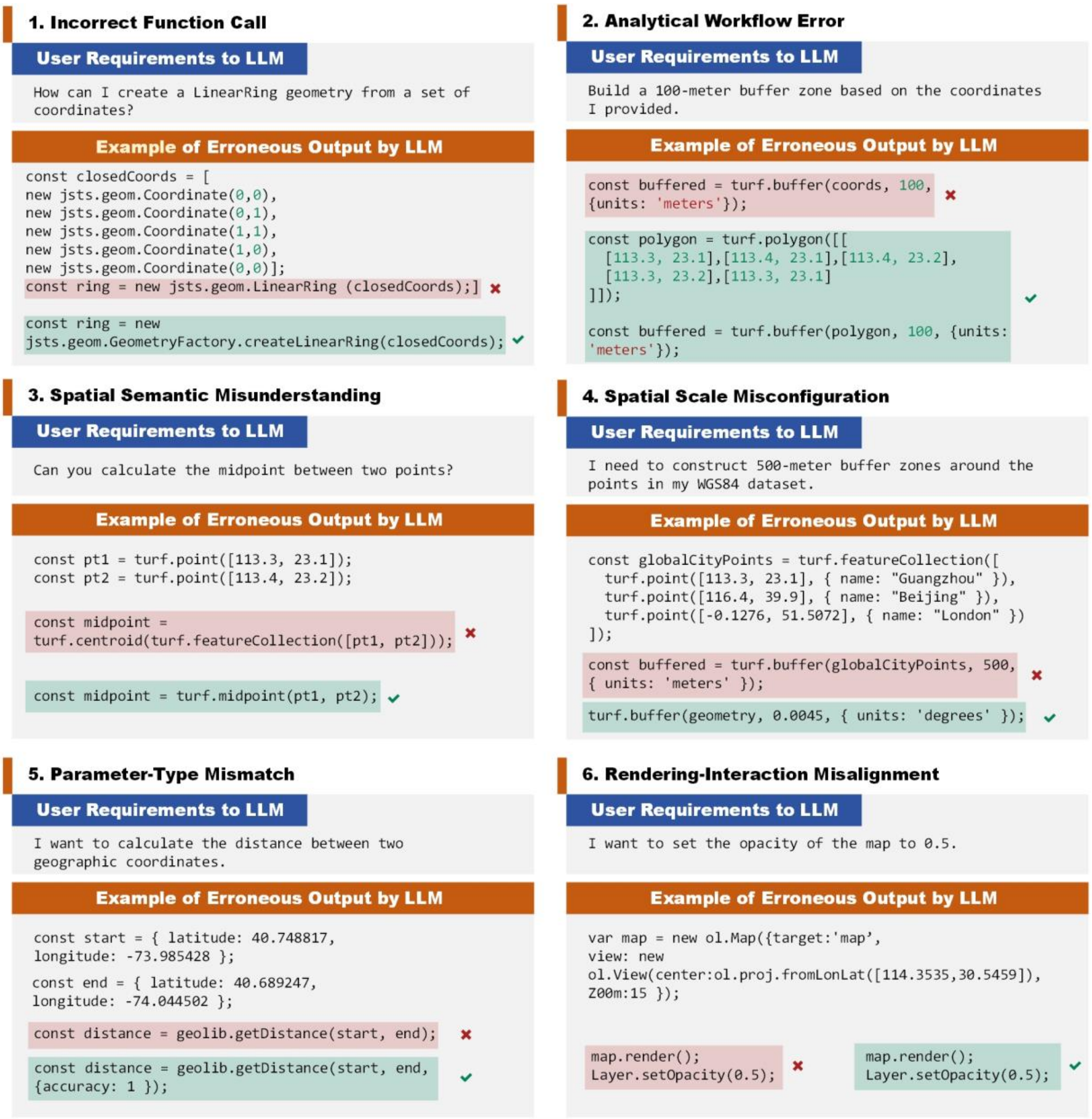

**Figure 1** Common error types in geospatial code generation with LLMs.

To address the aforementioned challenges, this study proposes and implements GeoJSEval, a LLM evaluation framework designed for unit-level function generation assessment. GeoJSEval establishes an automated and structured evaluation system for JavaScript-based geospatial code. As illustrated in **Figure 2**, GeoJSEval comprises three core components: the GeoJSEval-Bench test suite, the Submission Program, and the Judge Program. The GeoJSEval-Bench module (**Figure 2a**) is constructed based on the official documentation of five mainstream JavaScript geospatial libraries: Turf.js, JSTS, Geolib, Leaflet, and OpenLayers. The benchmark contains a total of 432 function-level tasks and 2,071 structured test cases. All test cases were automatically generated using multi-round prompting strategies with Claude Sonnet 4 and subsequently verified and validated by domain experts to ensure correctness and executability. Each test case encompasses three major components—function semantics, runtime configuration, and evaluation logic—and includes seven key metadata fields: function_header, reference_code, parameters_list, output_type, eval_methods, output_path, and expected_answer. The benchmark spans 25 data types supported across the five libraries, including geometric objects, map elements, numerical values, and boolean types—demonstrating high representativeness and diversity. The Submission Program (**Figure 2b**) uses the function header as a core prompt to guide the evaluated LLM in generating a complete function implementation. The system then injects predefined test parameters and executes the generated code, saving the output to a designated path for downstream evaluation. The Judge Program (**Figure 2c**) automatically selects the appropriate comparison strategy based on the output type, enabling systematic evaluation of code correctness and executability. GeoJSEval also integrates resource and performance monitoring modules to track token usage, response time, and code length, allowing the derivation of runtime metrics such as Inference Efficiency, Token Efficiency, and Code Efficiency. Additionally, the framework incorporates an error detection mechanism that automatically identifies common failure patterns—including syntax errors, type mismatches, attribute errors, and runtime exceptions—providing quantitative insights into model performance. In the experimental evaluation, we systematically assess a total of 18 state-of-the-art LLMs: 9 general non-reasoning models, 3 reasoning-augmented models, 5 code-centric models, and 1 geospatial-specialized model, GeoCode-GPT(Hou, Shen, Zhao, et al., 2025). The results reveal significant performance disparities, structural limitations, and potential optimization directions in the ability of current LLMs to generate JavaScript geospatial code.

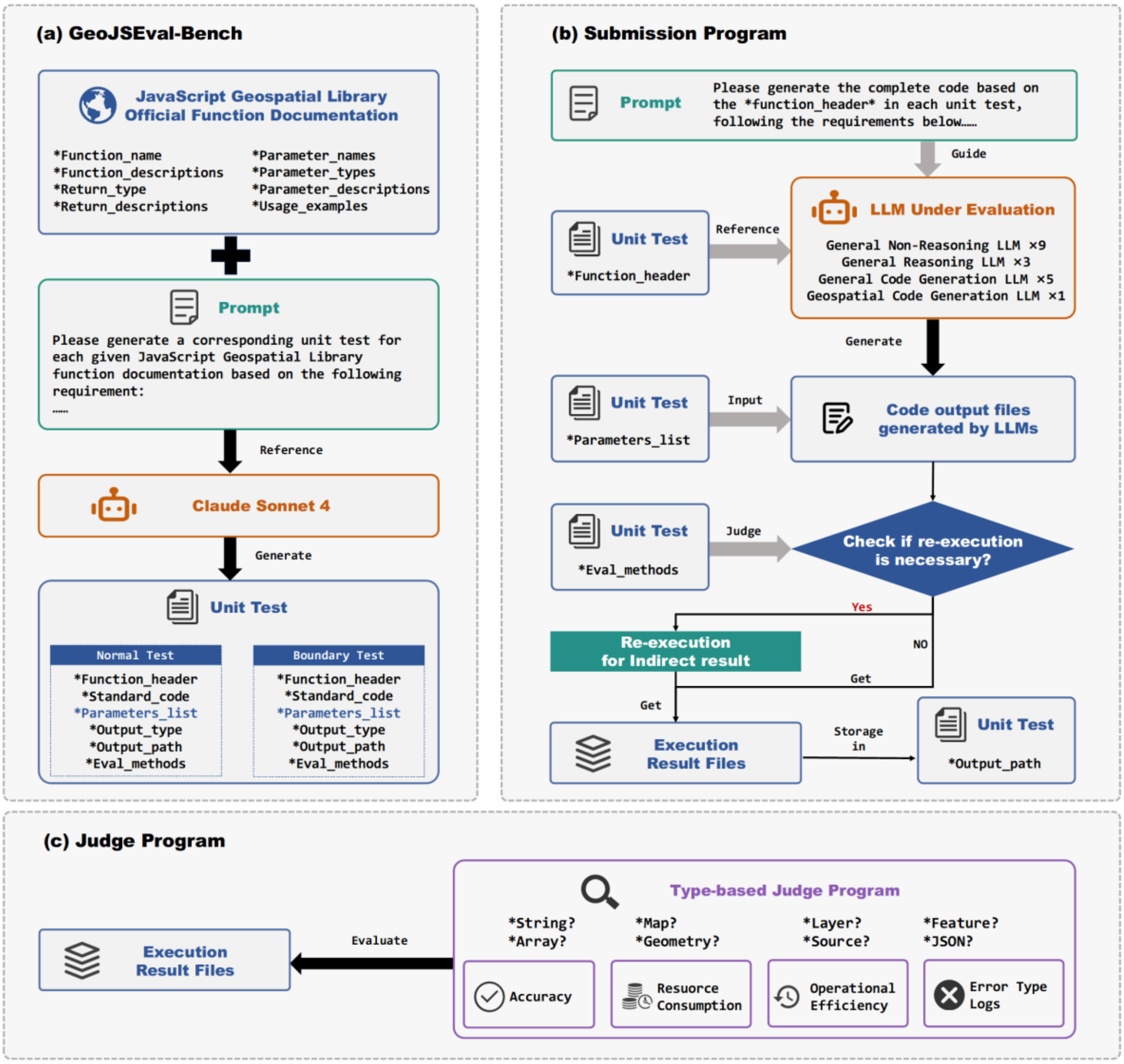

**Figure 2**. GeoJSEval framework structure. The diagram highlights GeoJSEval-Bench (a), the Submission Program (b), and the Judge Program (c).

The main contributions of this study are summarized as follows:

- We propose GeoJSEval, the first automated evaluation framework for assessing LLMs in JavaScript-based geospatial code generation. The framework enables a fully automated, structured, and extensible end-to-end pipeline—from prompt input and code generation to execution, result comparison, and error diagnosis—addressing the current lack of standardized evaluation tools in this domain.
- We construct and open-source GeoJSEval-Bench, a comprehensive benchmark consisting of 432 function-level test tasks and 2,071 parameterized cases, covering 25 types of JavaScript-based geospatial data. The benchmark systematically addresses core spatial analysis functions, enabling robust assessment of LLM performance in geospatial code generation.
- We conduct a systematic evaluation and comparative analysis of 18 representative LLMs across four model families. The evaluation quantifies performance across multiple dimensions—including execution pass rates, accuracy, resource consumption, and error type distribution—revealing critical bottlenecks in spatial semantic understanding and frontend ecosystem adaptation, and offering insights for future model optimization.

The remainder of this paper is organized as follows: Section 2 reviews related work on geospatial code, code generation tasks, and LLM-based code evaluation. Section 3 presents the task modeling, module design, construction method, and construction results of the GeoJSEval-Bench test suite. Section 4 details the overall evaluation pipeline of the GeoJSEval framework, including the implementation of the submission and judge programs. Section 5 describes the evaluated models, experimental setup, and the multidimensional evaluation metrics. Section 6 presents a systematic analysis and multidimensional comparison of the evaluation results. Section 7 concludes the study by summarizing key contributions, discussing limitations, and outlining future research directions.

## 2. Related Work

### 2.1 Geospatial Code

Geospatial code refers to the programmatic representation of logic for reading, processing, analyzing, and visualizing geospatial data. It plays a central role in enabling task automation and reusable modeling in modern Geographic Information Systems (GIS)(Li & Ning, 2023). Since the 1980s, geospatial code has undergone a multi-phase evolution—from embedded scripting in proprietary platforms, to cross-platform function libraries, and more recently to cloud computing and browser-based interactive environments. In the early stages, geospatial code was embedded within closed GIS platforms such as Arc/INFO, MapInfo, and GRASS GIS. Users performed basic tasks—such as vector editing, raster reclassification, and buffer generation—via command-line interfaces or macro languages. These systems were platform-dependent and lacked portability and extensibility. Since the 2000s, driven by the rise of open-source ecosystems, GIS programming has shifted toward function library paradigms built on general-purpose languages such as Python, R, and Java. Toolchains such as GDAL/OGR, Shapely, PySAL, and PostGIS have enhanced standardization and cross-platform operability in geospatial analysis workflows(Pakdil & Çelik, 2022). In parallel, mainstream desktop GIS platforms introduced scripting APIs—such as ArcPy for ArcGIS and PyQGIS for QGIS—enabling users to build plugins and automated workflows for customizable spatial

analysis(Hu et al., 2025). Since the 2010s, the advent of cloud computing has driven the evolution of geospatial code toward service-oriented architectures. Platforms like GEE offer cloud-based processing of global-scale remote sensing and spatiotemporal data through JavaScript and Python APIs, enabling a "Code-as-a-Service" paradigm for geospatial modeling. With the proliferation of WebGIS applications, JavaScript has emerged as a key language due to its native browser execution and high interactivity. Frontend libraries such as Leaflet, OpenLayers, and Mapbox support map rendering and interactive control, while analytical libraries like Turf.js, JSTS, and Geolib enable buffer analysis, geometric operations, and spatial relationship evaluation. These tools collectively foster a transition from backend geospatial coding to a "computable web" architecture(Gui et al., 2013). JavaScript's lightweight nature, cross-platform compatibility, and client-side computing capabilities position it as a powerful medium for real-time visualization and interactive spatial analysis.

### 2.2 Code Generation

Code generation originated in the general-purpose programming domain and has evolved through three major phases: rule-based, data-driven, and LLM-based approaches. Since the 1980s, early code generation techniques primarily relied on manually defined rules and templates to automate code creation. For example, visual programming languages enabled users to generate code by dragging and dropping graphical modules. While suitable for educational or low-complexity tasks, such systems lacked the capacity to model semantic and logical dependencies, limiting their applicability in complex software development(Burnett et al., 1995). Beginning around 2010, the increasing availability of large-scale code datasets and advancements in deep learning enabled the emergence of data-driven code generation approaches. These methods learned from existing code samples to perform tasks such as structure completion and syntax correction. Systems like DeepCode, for instance, learned general coding patterns from open-source repositories to support code completion and provide security suggestions(Qu et al., 2024). However, these approaches often relied on static feature extraction, exhibited limited generalization capabilities, and were heavily dependent on annotated training data. Since 2020, the advent of LLMs has ushered in a new era of code generation. CodeBERT introduced a dual-modality pretraining strategy that jointly learns from natural and programming languages, supporting tasks such as code completion and Natural Language to Code (NL2Code) generation(Xu et al., 2022). Subsequently, models such as OpenAI Codex and PolyCoder leveraged large-scale natural language–code pair datasets to enable end-to-end automation from requirement understanding to multilingual code generation, thereby establishing the NL2Code paradigm. Models developed during this stage not only demonstrated strong language comprehension, but also began to exhibit emerging capabilities in code reasoning and task planning, leading to widespread deployment in real-world software development tools(Storey, 2006).

Compared to general-purpose code generation, research on geospatial code generation emerged relatively late. Early efforts primarily relied on platform-embedded template systems and rule-based engines. Examples include the parameterized scripting wizards in GEE, the ModelBuilder in Esri ArcGIS Pro, and subsequent AI-assisted suggestion tools, all of which focused on improving function call efficiency but lacked support for cross-platform generalization and natural language task interpretation. It was not until October 2024, with the publication of two seminal benchmark studies, that geospatial code generation was formally defined as an independent research task. These works extended the general NL2Code paradigm into NL2GeospatialCode, providing the theoretical foundation for this emerging field. Since then, the field has experienced rapid growth, with a proliferation of optimization strategies. For example, the Chain-of-Programming (CoP) strategy enables task decomposition via programming thought chains(Hou, Jiao, et al., 2025); Geo-FuB and GEE-Ops construct functional semantic and API knowledge bases, which, when integrated with Retrieval-Augmented Generation (RAG), significantly enhance code generation accuracy(Hou et al.; Hou, Zhao, et al., 2025). GeoCode-GPT further advances the field by fine-tuning LLMs on geoscientific code corpora, making it the first domain-specialized LLM for geospatial code generation(Hou, Shen, Zhao, et al., 2025). In addition, agent-based frameworks such as GeoAgent, GIS Copilot, and ShapefileGPT incorporate task planners, tool callers, and feedback loops to support end-to-end workflows from natural language understanding to geospatial code generation and autonomous execution—forming the preliminary blueprint of a "self-driving GIS" system(Gwak et al., 2019). Collectively, these advancements signify a shift of geospatial code generation from niche applications toward mainstream research in system modeling and model optimization.

Nevertheless, significant challenges remain, including strong platform dependency, complex function structures, heterogeneous data formats, and high semantic reasoning difficulty(Hochmair et al., 2025; Palomino et al., 2017). Furthermore, the absence of standardized evaluation protocols and reproducible testbeds limits the scalability of model development and impedes fair cross-model comparison. Therefore, establishing a comprehensive evaluation framework for geospatial code generation—one that supports multi-platform, cross-language, and task-complete assessments—remains a critical step toward advancing this field.

### 2.3 Evaluation of Code Generation

Evaluation frameworks for code generation tasks have undergone a continuous evolution from syntax validation to semantic understanding. Early approaches primarily focused on syntactic correctness, functional executability, and output consistency—typically evaluated by directly comparing the generated output against the expected results. For example, Sutherland's 1988 theory on code generation correctness emphasized syntactic precision and logical consistency, yet lacked consideration for multidimensional aspects of code quality(Benveniste et al., 2000). Since 2010, with the emergence of large-scale annotated datasets and automated verification systems, evaluation has progressively shifted toward a data-driven paradigm. For instance, Google conducted comprehensive evaluations of Python code generation within neural machine translation frameworks, covering aspects such as functional correctness, execution efficiency, and structural soundness—substantially enhancing both accuracy and applicability of evaluation. Since 2020, the advancement of LLMs has led to the standardization of evaluation benchmarks. Datasets such as HumanEval, MBPP, and APPS are now widely adopted to assess code executability, semantic fidelity, and task generalization(Le-Cong et al., 2023). Evaluation protocols proposed by OpenAI further broaden the scope by incorporating deep understanding of complex code structures and logical reasoning, with a particular emphasis on models' capabilities in semantic comprehension and intent recognition(Wei et al., 2024).

Compared to general-purpose code, evaluating geospatial code generation presents substantially more complex challenges, primarily due to the difficulty of handling multimodal heterogeneous data and the strong dependency on specific platforms. Geospatial data spans various formats, including vector and raster types.

However, platforms differ significantly in terms of data structures, access mechanisms, and operator invocation logic, often resulting in semantic inconsistencies when generated code is executed across different environments. Against this backdrop, evaluating geospatial code in the JavaScript environment introduces additional dimensions of complexity. In recent years, JavaScript has been widely adopted in geospatial visualization and analysis scenarios, owing to its native browser execution and high interactivity. Browser-executable geospatial code is characterized by lightweight design, asynchronous execution, and tightly coupled logic. Evaluating such code requires not only verifying syntactic and logical correctness, but also assessing spatial semantic comprehension, API interface alignment, browser runtime behavior, and interaction fidelity. However, JavaScript-based geospatial code remains largely unaddressed in mainstream code evaluation benchmarks, and lacks dedicated, systematic assessment frameworks. Some preliminary efforts have been made to evaluate geospatial code generation—for example, Wuhan University proposed GeoCode-Bench and GeoCode-Eval—but these rely heavily on manual expert review, leading to issues of subjectivity and poor reproducibility. The University of Wisconsin conducted a preliminary evaluation of GPT-4 for ArcPy code generation, yet the associated data and methodology were not publicly released, and the scope of task complexity and platform diversity was limited. AutoGEEval and its extension AutoGEEval++ introduced automated evaluation pipelines for the GEE platform, improving systematization and execution validation. Nevertheless, their support for JavaScript frontend ecosystems and diverse task types remains limited(Hou, Shen, Wu, et al., 2025; Wu et al., 2025).

## 3. GeoJSEval-Bench

This study presents GeoJSEval-Bench, a standardized benchmark suite designed for unit-level evaluation of geospatial function calls. It aims to assess the capability of LLMs in generating JavaScript-based geospatial code. The benchmark is constructed based on the official API documentation of five representative open-source JavaScript geospatial libraries, covering core task modules ranging from spatial computation to map visualization. These libraries include Turf.js, JSTS, Geolib, Leaflet, and OpenLayers. These libraries can be broadly categorized into two functional groups. The first group focuses on spatial analysis and geometric computation. Turf.js offers essential spatial analysis functions such as buffering, union, clipping, and distance calculation, effectively covering core geoprocessing operations in frontend environments. JSTS specializes in topologically correct geometric operations, supporting high-precision analysis for complex geometries. Geolib focuses on point-level operations in latitude - longitude space, providing precise distance calculation, azimuth determination, and coordinate transformation. It is particularly valuable in location-based services and trajectory analysis. The second group comprises libraries for map visualization and user interaction. Leaflet, a lightweight web mapping framework, offers core APIs for map loading, tile layer control, event handling, and marker rendering. It is widely used in developing small to medium-scale interactive web maps. In contrast, OpenLayers provides more comprehensive functionalities, including support for various map service protocols, coordinate projection transformations, vector styling, and customizable interaction logic—making it a key tool for building advanced WebGIS clients. Based on the function systems of these libraries, GeoJSEval-Bench contains 432 unit-level test cases, spanning 25 data types, including map objects, layer objects, strings, and numerical values. This chapter provides a detailed description of the unit-level task modeling, module design, construction methodology, and the final composition of the benchmark.

### 3.1 Task Modeling

The unit-level testing task ($\mathcal{T}_{unit}$) proposed in GeoJSEval is designed to perform fine-grained evaluation of LLMs in terms of their ability to understand and generate function calls for each API function defined in five representative JavaScript geospatial libraries. This includes assessing their comprehension of function semantics, parameter structure, and input-output specifications. Specifically, each task evaluates whether a model can, given a known function header, accurately reconstruct the intended call semantics, generate a valid parameter structure, and synthesize a syntactically correct, semantically aligned, and executable JavaScript code snippet $C$ that effectively invokes the target function $f$. This process simulates a typical developer workflow—reading API documentation and writing invocation code—and comprehensively assesses the model's capability chain from function comprehension to code construction and finally behavioral execution. Fundamentally, this task examines the model's code generation ability at the finest operational granularity—a single function call—serving as a foundational measure of an LLM's capacity for function-level behavioral modeling.

Let the set of functions provided in the official documentation of the five mainstream JavaScript geospatial libraries be denoted as:

$$\mathcal{F} = \{f_1, f_2, \dots, f_N\},\ f_i \in GeoJS_API \tag{1}$$

For each model under evaluation, the task is to generate a code snippet $C_i$ that conforms to JavaScript syntax specifications and can be successfully executed in a real runtime environment:

$$\mathcal{T}_{unit}: f_i \rightarrow C_i \tag{2}$$

GeoJSEval formalizes each function-level test task as the following function mapping process:

$$\mathcal{M}(H) \Rightarrow C \Rightarrow y = Exec(C, f_i) \Rightarrow Judge(y, a) \tag{3}$$

Specifically, $H$ denotes the function header, including the function name and parameter signature. $\mathcal{M}$ refers to the large language model, which takes $H$ as input and produces a function call code snippet $C$. The symbol $f_i$ represents the target test function, and $y$ denotes the execution result of the generated code $C$ applied to $f_i$. The expected reference output is denoted by $a$. The module Exec represents the code execution engine, while Judge is the evaluation module responsible for determining whether $y$ and $a$ are consistent. Furthermore, GeoJSEval defines a task consistency function $L(q)$ to assess whether the model behavior aligns with expectations over a test set $Q$. The equality operator "=" used in this comparison may represent strict equivalence, floating-point tolerance, or set-based inclusion—depending on the semantic consistency requirements of the task.

$$L(q) = \begin{cases} 1, & if\ E(Exec(\mathcal{M}(H), i), a) = True \\ 0, & otherwise \end{cases} \tag{4}$$

### 3.2 Module Design

All unit-level test cases were generated based on reference data using Claude Sonnet 4, an advanced large language model developed by Anthropic, guided by predefined prompts. Each test case was manually reviewed and validated by domain experts to ensure correctness and executability (see Section 3.3 for details).

Each complete test case is structured into three distinct modules: the Function Semantics Module, the Execution Configuration Module, and the Evaluation Parsing Module.

### 3.2.1 Function Semantics Module

In the task of automatic geospatial function generation, accurately understanding the invocation semantics and target behavior of a function is fundamental to enabling the model to generate executable code correctly. Serving as the semantic entry point of the entire evaluation pipeline, the Function Semantics Module aims to clarify two key aspects: (i) what the target function is intended to accomplish, and (ii) how it should be correctly implemented. This module comprises two components: the function header, which provides the semantic constraints for input, and the standard_code, which serves as the behavioral reference for correct implementation.

- The function declaration $\mathcal{H}_i \in$ FunctionHeader provides the structured input for each test task. It includes the function name, parameter types, and semantic descriptions. This information is combined with prompt templates to offer semantic guidance for the large language model, steering it to generate the complete function body. The structure of $\mathcal{H}_i$ is formally defined as:

$$\mathcal{H}_i = (name, params, types, desc) \tag{5}$$

  Here, name denotes the function name; $\text{params} = \{p_1, p_2, \dots, p_n\}$ represents the list of input parameters; $\text{types}(p_i)$ specifies the type information for each parameter $p_i$; and "desc" provides a natural language description of the function's purpose and usage scenario.

- The reference code snippet $\mathcal{S}_i \in$ StandardCode provides the correct implementation of the target function under standard semantics. It is generated by Claude Sonnet 4 using a predefined prompt and is executed by domain experts to produce the ground-truth output. The behavior of this function should remain stable and reproducible across all valid inputs, that is:

$$x = \mathcal{S}_i(params), \textit{for all valid input params} \tag{6}$$

  The resulting output $x$ serves as the reference answer $a$ in the subsequent evaluation logic module. It is important to note that this reference implementation is completely hidden from the model during evaluation. The model generates the test function body solely based on the function declaration $\mathcal{H}_i$, without access to $\mathcal{S}_i$.

### 3.2.2 Execution Configuration Module

GeoJSEval incorporates an Execution Configuration Module into each test case, which explicitly defines all prerequisites necessary for execution in a structured manner. This includes parameters_list, output_type, edge_test, and output_path. Together, these elements ensure the completeness and reproducibility of the test process.

- The parameter list $\mathcal{P}_i \in$ ParameterList explicitly specifies the concrete input values required for each test invocation. These structured parameters are injected as fixed inputs into both the reference implementation $S_i$ and the model-generated code $C_i$, enabling the construction of a complete and executable function invocation and driving the generation of output results.
- The output type $\mathcal{T}_i \in \text{OutputType}$ specifies the expected return type of the function and serves as

a constraint on the format of the model-generated output. Upon executing the model-generated code and obtaining the output $y$, the system first verifies whether its return type matches the expected type $\mathcal{T}$. If a type mismatch is detected, the downstream consistency evaluation is skipped to avoid false judgments or error propagation.

- The edge test item $\mathrm{E}_i \in \mathrm{EdgeTest}$ indicates whether the boundary testing mechanism is enabled for this test case. If the value is not empty, the system will automatically generate a set of extreme or special boundary parameters based on the standard input values and inject them into the test process. This is done to assess the model's robustness and generalization ability under extreme input conditions.
- The output path $\mathcal{O}_i \in \mathrm{OutputPath}$ specifies the storage location of the output results produced by the model during execution. The testing system will read the external result file produced by the model from the specified path.

### 3.2.3 Evaluation Parsing Module

The evaluation logic module is responsible for defining the decision-making markers that the testing framework will use during the evaluation phase. It clarifies whether the results of the model-generated code are comparable, whether structural information needs to be extracted, and whether auxiliary functions are required for comparison. Serving as the input for the judging logic, this module consists of two components: expected_answer and eval_methods. Its core responsibility is to provide a structured, automated mechanism for evaluability annotation.

- The expected answer $\mathcal{A}_i \in \mathrm{ExpectedAnswer}$ refers to the correct result obtained by executing the reference code snippet with the predefined parameters, serving as the sole benchmark for comparison during the evaluation phase, used to assess the accuracy of the model's output. The generation process is detailed in Section 3.2.1, which covers the Function Semantics Module.
- The evaluation method $\mathcal{E}_i \in \mathrm{EvalMethods}$ indicates whether the function output can be directly evaluated. If direct evaluation is not possible, it specifies the auxiliary function(s) to be invoked. This field includes two types: direct evaluation and indirect evaluation. Direct evaluation applies when the function output is a numerical value, boolean, text, or a parseable data structure (e.g., JSON, GeoJSON), which can be directly compared. For such functions, $\mathcal{E}_i$ is marked as null, indicating that no further processing is required. Indirect evaluation applies when the function output is a complex object (e.g., Map, Layer, Source), requiring the use of specific methods from external libraries to extract relevant features. In this case, $\mathcal{E}_i$ explicitly marks the required auxiliary functions for feature extraction. For auxiliary function calls, based on the function's role and return type, we select functions from the platform's official API that can extract indirect semantic features of complex objects (e.g., map center coordinates, layer visibility, feature count in a data source). This ensures accurate comparison under structural consistency and semantic alignment. This paper constructs an auxiliary function library for three typical complex objects in geospatial visualization platforms (Map, Layer, Source), as shown in **Table 1**. Depending on the complexity of the output structure of different evaluation functions, a single method or a combination of auxiliary methods may be chosen to ensure the accuracy of high-level semantic evaluations.

**Table 1**. Auxiliary Evaluation Methods for Indirect Output Objects

| Object Type | Functionality | Auxiliary Method |
|---|---|---|
| **Map** | Get center coordinates | map.getCenter() |
| | Get zoom level | map.getZoom() |
| | Get resolution | map.getResolution() |
| | Get rotation angle | map.getRotation() |
| | Get map extent | map.getExtent() |
| | Get number of layers | map.getLayers().getLength() |
| | Get layer collection | map.getLayers() |
| **Layer** | Get opacity | layer.getOpacity() |
| | Check visibility | layer.getVisible() |
| | Get layer extent | layer.getExtent() |
| | Get z-index | layer.getZIndex() |
| | Get layer zoom level | layer.getZoom() |
| | Get data source object | layer.getSource() |
| **Source** | Get all features | source.getFeatures() |
| | Query feature by ID | source.getFeatureById(id) |
| | Get number of features | source.getFeatureCount() |

### 3.3 Construction Method

Based on task modeling principles and modular design concepts, GeoJSEval-Bench has constructed a clearly structured and automatically executable unit test benchmark, designed for the systematic evaluation of LLMs in geospatial function generation tasks. This benchmark is based on the official API documentation of five prominent JavaScript geospatial libraries—Turf.js, JSTS, Geolib, Leaflet, and OpenLayers—covering a total of 432 function interfaces. Each function page provides key information, including the function name, functional description, parameter list with type constraints, return value type, and semantic explanations. Some functions come with invocation examples, while others lack examples and require inference of the call structure based on available metadata. **Figure 3** illustrates a typical function documentation page.

During the construction process, the research team first conducted a structured extraction of the API documentation for the five target libraries, standardizing the function page contents into a unified JSON format. Based on this structured representation, the framework designed dedicated Prompt templates for each function to guide the large language model in reading the semantic structure and generating standardized test tasks. The template design carefully considers the differences across libraries in function naming conventions, parameter styles, and invocation semantics. **Figure 4** presents a sample Prompt for a function from the OpenLayers library. Each automatically generated test task is subject to a strict manual review process. The review process is conducted by five experts with geospatial development experience and proficiency in JavaScript. The review focuses on whether the task objectives are reasonable, descriptions are clear, parameters are standardized, and whether the code can execute stably and return logically correct results. For tasks with runtime anomalies, unclear semantics, or structural issues, experts make necessary revisions based on their professional judgment. The background information and selection criteria for the expert team are detailed in **Table 2**. All tasks that pass the review and execute successfully have their standard outputs saved in the specified output_path as the reference answer. During the testing phase, the judging program automatically reads the reference results from this path and compares them with the model-generated output, serving as the basis for evaluating accuracy.

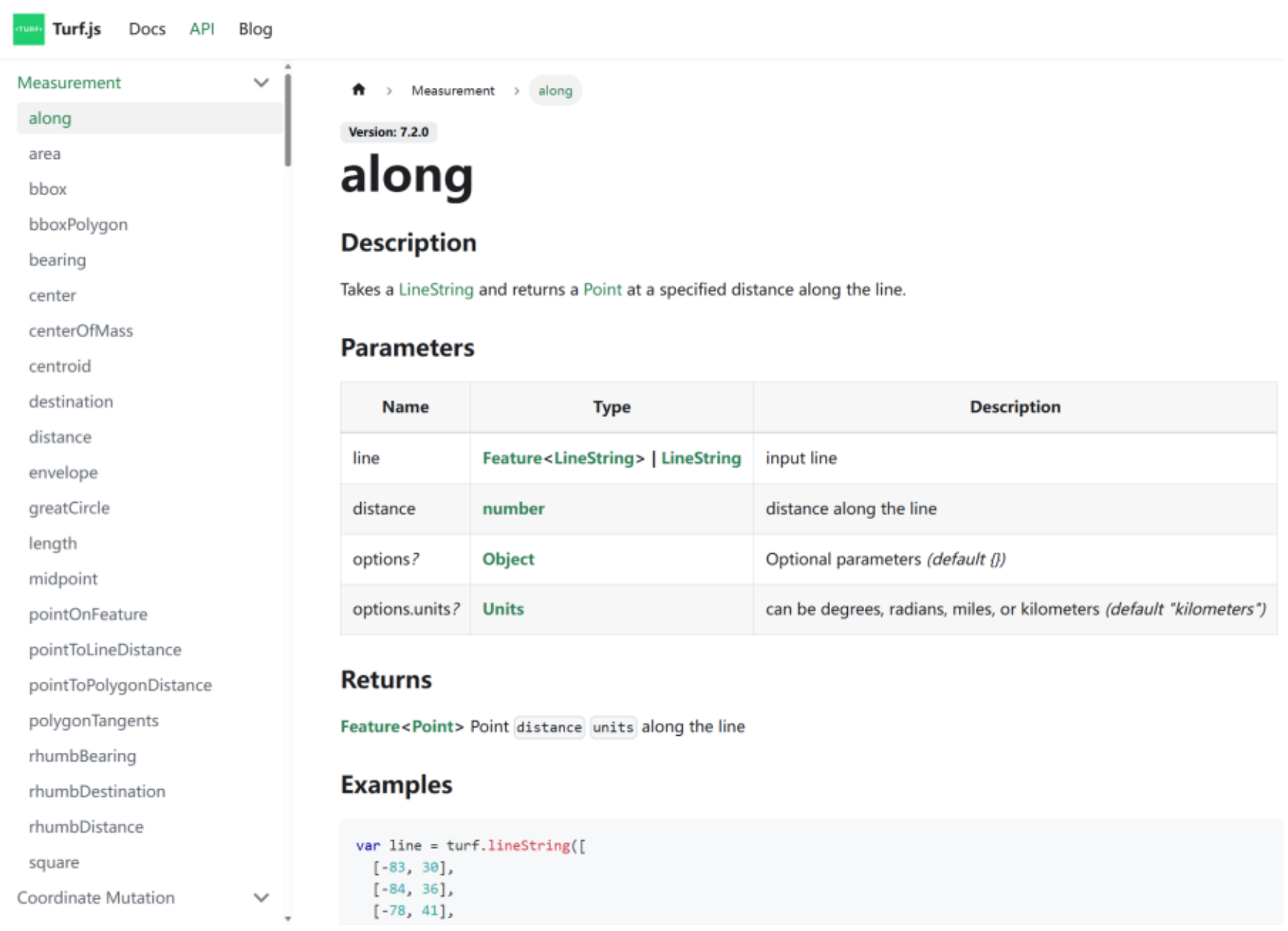

**Figure 3.** Example Page from Turf.js API Reference.

**Prompt**

```
###  Task Description:

You are an advanced JavaScript developer with expertise in the OpenLayers library. This task consists of two stages: **function encapsulation** and **automated YAML test configuration**. The requirements are as follows:

###  Stage 1: Encapsulation Function Development

Based on a pre-initialized OpenLayers map (including `map`, `osmLayer`, and `view`), encapsulate a new function prefixed with `O_` using a given built-in function. Requirements:

1. **Descriptive function names**, e.g., `O_clonePointGeometry`;
2. **Only use required parameters**; optional parameters should be fixed internally;
3. **If the original function has no return value**, return the affected object (`map`, `view`, or `layer`);
4. **Keep function implementation simple**, directly calling the built-in function;
5. **Add inline JSDoc-style comments in English**, including `@param` and `@return`, with two fixed introductory sentences;
6. **Provide 5 test cases**: 3 normal and 2 boundary cases. Examples should include only parameter definitions and function calls, reuse predefined objects like `map` or `osmLayer`, and avoid using `console.log`.

###  Stage 2: YAML-Based Evaluation Configuration

Generate a YAML configuration for the encapsulated function, enabling automated test scripts to load parameter definitions and evaluation methods. Structure details:

####  YAML Field Specification (fixed order):

* `params`: Dictionary of function parameters

  * Direct assignment, e.g., `zoom: 15`
  * Or via `!js |` with JavaScript function, e.g.:

    ```yaml
    point: !js |
      function createPoint() {
        return new ol.geom.Point(...);
      }
    ```

  * Use `map: true` to reference pre-initialized objects
  * If modified, define using `!js` with full reconstruction and return

* `eval_methods`: Evaluation expressions

  * `[]` for simple return types
  * For objects, provide expressions like `["map.getView().getZoom()"]`

* `out_type`: Output type, e.g., `"ol.View"` or `"Boolean"`

* `expected_answer`: File name of expected result, e.g., `O_setMapViewZoom_1_false.json`

* `edge_test`: Boolean indicating boundary case

* `Which`: Constant value `"OpenLayers"`

•  `Visual`: Constant value `true`

•  The function information to be processed is as follows:
```

**Figure 4.** Prompt for test construction.

**Table 2.** Background and Selection Criteria of Experts

| No. | Age | Qualification | Selection Criteria |
|---|---|---|---|
| 1 | 45 | Senior Professor | A leading scholar in the field of geospatial information science with in-depth expertise in remote sensing data processing and spatial data analysis. This expert has authored over 100 academic publications, led several national-level research projects, and contributed significantly to the development of spatial modeling techniques and geospatial computing frameworks. |
| 2 | 38 | Professor | Specializes in geospatial data mining, spatiotemporal data analysis, and modeling, with a particular emphasis on the integration of heterogeneous spatial datasets. Has published extensively in high-impact international journals in collaboration with institutions such as NASA and USGS. Research focuses on transportation network optimization, trajectory mining, and scalable computation techniques for processing large-scale geospatial datasets across urban and environmental domains. |
| 3 | 40 | Senior Algorithm Engineer | Senior algorithm engineer with over 15 years of experience in artificial intelligence and natural language processing, with a strong focus on the development of programming language models and automatic code generation systems. Specializes in large-scale pretrained language models tailored for software engineering tasks, including syntax parsing, code translation, and multi-language code synthesis. |
| 4 | 50 | Professor | Professor specializing in geospatial visualization and web-based mapping technologies, with extensive expertise in ecological and agricultural geospatial data analysis. Experienced in building interactive mapping platforms using libraries such as OpenLayers and Cesium. His research focuses on integrating spatial statistics, remote sensing, and web GIS for land monitoring and decision-making support. |
| 5 | 28 | PhD Candidate | PhD candidate researching the integration of geospatial analytics and automated code generation. With a solid foundation in computer science and GIS, she focuses on improving clarity, usability, and efficiency in spatial development workflows. Her work explores how intelligent code tools can support transparent, accessible, and robust geospatial modeling and application development. |

### 3.4 Construction Results

GeoJSEval-Bench ultimately constructed 432 primary function-level test tasks, covering all well-defined and valid functions across the five target libraries. Based on this foundation, and incorporating various input parameters and invocation path configurations, an additional 2,071 structured unit test cases were generated to systematically cover diverse use cases. The test tasks encompass a wide variety of data types, reflecting common data structures in geospatial development practice. A total of 25 types are covered, with the detailed list provided in **Table 3**. From an evaluability perspective, these data types can be broadly categorized into two groups: one consists of basic types with clear structures and stable semantics, such as numeric variables, strings, arrays, and standard GeoJSON. These types of data can be directly parsed into structured results and are suitable for direct comparison in evaluation. The other group consists of more complex encapsulated object types, such as Map and Layer objects in Leaflet or OpenLayers, where internal state information cannot be accessed directly and requires the use of auxiliary methods to extract semantic information and align the structure. To aid in understanding the differences between these types and their corresponding evaluation strategies, **Figure 5** shows a direct evaluation test case example, while **Figure 6** demonstrates a test case for complex object types requiring indirect structural extraction. A few examples are provided here, with additional detailed examples available in Appendix A.

**Table 3**. Output Types in GeoJSEval-Bench

| **Source_library** | **Output_type** | **Description** |
|---|---|---|
| Standard JavaScript types | Array | Ordered list of numbers and pixels. |
| | Boolean | True or false value. |
| | String | Textual data type. |
| | Number | Numeric value (int or float). |
| Standard geometry types | Geometry | Geometric shapes (point, line, polygon, etc.) |
| | GeometryCollection | Group of multiple geometries. |
| | Feature | Geometry with properties. |
| | FeatureCollection | Set of multiple features. |
| Geolib-specific | geolib.coordinates | Coordinate array for distance calculation. |
| | geolib.distanceCoordinate | Coordinate with distance metadata. |
| | geolib.bounds | Geographic bounding box. |
| | geolib.center | Geometric center of coordinates. |
| Leaflet-specific | leaflet.Map | Main Leaflet map object. |
| | leaflet.Layer | Visual overlay on the map. |
| | leaflet.Latlng | Geographic point (lat, lng). |
| | leaflet.LatlngBounds | Rectangular geographic extent. |
| | leaflet.Point | Pixel-based 2D point. |
| | leaflet.Bounds | Pixel bounds of an area. |
| Openlayers-specific | ol.Map | Main OpenLayers map object. |
| | ol.View | Map view (center, zoom, etc.). |
| | ol.Layer | Data rendering layer. |
| | ol.Source | Data provider for layers. |
| | ol.Coordinate | Geographic coordinate pair. |
| | ol.Extent | Bounding box [minX, minY, maxX, maxY]. |
| | ol.Size | Pixel dimensions [width, height]. |

**1 Number**

**Function_header**

```
function G_calculatePathLengthInMeters(points) {
    /**
     * The Geolib library has already been pre-loaded, so don't load it again.
     *
     * Use the given built-in functions from Geolib to complete the task as simply
as possible.
     *
     * Calculates the total length (in meters) of a path defined by a collection of
geographic coordinates. Uses geolib's default getDistance function for calculations.
     *
     * @param {Array<Object>} points - Array of coordinate objects in {latitude:
number, longitude: number} format
     * @return {number} Total path length in meters
     */
}
```

**Standard_code**

```
function G_calculatePathLengthInMeters(points) {
    /**
     * The Geolib library has already been pre-loaded, so don't load it again.
     *
     * Use the given built-in functions from Geolib to complete the task as simply
as possible.
     *
     * Calculates the total length (in meters) of a path defined by a collection of
geographic coordinates. Uses geolib's default getDistance function for calculations.
     *
     * @param {Array<Object>} points - Array of coordinate objects in {latitude:
number, longitude: number} format
     * @return {number} Total path length in meters
     */
    return geolib.getPathLength(points);
}
```

**Parameter_list**

```
- params:
      points: !js |
        function get_Points() {
          const cityPath = [
            { latitude: 40.7128, longitude: -74.0060 },
            { latitude: 34.0522, longitude: -118.2437 },
            { latitude: 41.8781, longitude: -87.6298 }
          ];
          return cityPath;
        }
```

| | |
|---|---|
| **Output_type** | number |
| **Eval_methods** | [] |
| **Output _path** | G_calculatePathLengthInMeters_testcase.npy |
| **Expected _answer** | 6782342 |

**2 Geometry**

**Function_header**

```
function T_envelope(GeoJSON) {
    /**
     * The Turf library has already been pre-loaded, so don't load it again.
     *
     * Use the given built-in functions from Turf to complete the task as simply as
possible.
     *
     * Takes the given GeoJSON features and returns the encompassing Polygon.
     *
     * @param {GeoJSON} GeoJSON - The GeoJSON object containing features (Point,
LineString, MultiPoint, etc.) to be enveloped.
     * @return {GeoJSON.Polygon} - The smallest polygon that can contain all input
features.
     */
}
```

**Standard_code**

```
function T_envelope(GeoJSON) {
    /**
     * The Turf library has already been pre-loaded, so don't load it again.
     *
     * Use the given built-in functions from Turf to complete the task as simply as
possible.
     *
     * Takes the given GeoJSON features and returns the encompassing Polygon.
     *
     * @param {GeoJSON} GeoJSON - The GeoJSON object containing features (Point,
LineString, MultiPoint, etc.) to be enveloped.
     * @return {GeoJSON.Polygon} - The smallest polygon that can contain all input
features.
     */
    return turf.envelope(GeoJSON);
}
```

**Parameter_list**

```
- params:
      GeoJSON:
        type: "Feature"
        geometry:
          type: "LineString"
          coordinates: [[0, 0], [5, 5]]
```

| | |
|---|---|
| **Output_type** | geometry |
| **Eval_methods** | [] |
| **Output _path** | T_envelope_testcase.geojson |

**Expected _answer**

```
{
  "type": "Feature",
  "geometry": {
    "type": "Polygon",
    "coordinates": [[[0, 0],[5, 0],[5, 5],[0, 5],[0, 0]]]
  },
  "properties": {}
}
```

**Figure 5.** Unit Test Example for Direct Evaluation of Primitive Output Types

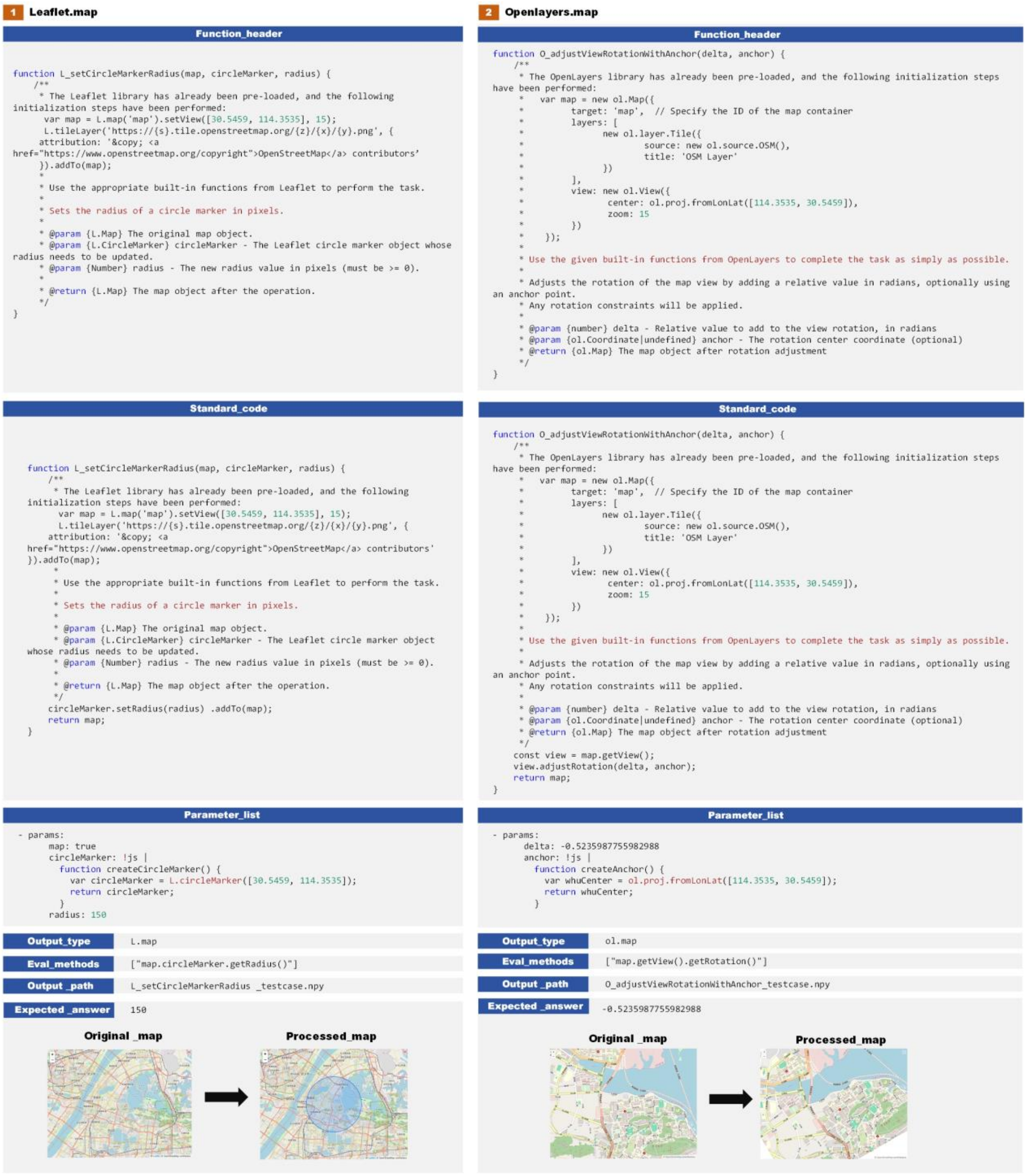

**Figure 6.** Unit Test Example for Indirect Evaluation of Complex Geospatial Objects

## 4. Submission and Judging Programs

During the evaluation phase, the GeoJSEval framework constructs an automated evaluation system through the collaboration of the Submission Program and the Judge Program, covering the entire process from code generation to result determination. The Submission Program is responsible for scheduling the large language model to generate and execute function call code based on the GeoJSEval-Bench test suite. The Judge Program systematically evaluates the model's output by comparing it with the reference answer. It applies various equivalence determination strategies based on output type and semantic features. This chapter outlines the core design and operational flow of the Submission Program and Judge Program.

### 4.1 Submission Program

The Submission Program is responsible for scheduling the evaluation of each test task in GeoJSEval-Bench by the large language model. As shown in **Figure 2b**, the overall process includes four key modules: code generation, evaluability determination, code execution, and result storage. During the generation phase, the system uses the structured function header provided by the Function Semantics Module as the sole input, combined with predefined prompt templates, to guide the model in generating the target function body. The prompt explicitly instructs the model to output only the JavaScript function body, without any additional text or explanations (as shown in **Figure 7**), ensuring that the generated code is executable and free from unstructured interference.

**Prompt**

````
prompt = (
            "Please complete the following JavaScript function for local geospatial
processing, based on the function signature and the provided JSDoc-style docstring. "
            "Return ONLY the complete function code without any explanations or
additional text. "
            "Do not add any comments beyond what's already in the docstring. Keep the
given docstring. "
            "The input may be empty or invalid, you need to handle this situation
appropriately to avoid program crashes. "
            "Wrap the code with ```javascript and ``` to indicate the code block.\n\n"
            "Here is the function:\n\n"
            f "{test_content}"
        )
````

**Figure 7.** Prompt for submission program

After the model generates the code, the system injects input values into the function call based on the parameters_list set in the Execution Configuration Module. The structured parameter values are mapped to the formal parameter positions in the model-generated JavaScript code, constructing a complete executable function call expression. This ensures that the generated code has all the necessary contextual information for execution in a uniform environment. Subsequently, the system refers to the eval_methods field in the Evaluation Logic Module to determine how the output should be stored and whether it needs to enter an auxiliary processing path. If eval_methods is empty, it indicates that the function output is a structured basic type (e.g., numeric, boolean, array, GeoJSON), which can be directly obtained through code execution. The result is then saved to the specified output_path. If eval_methods specifies auxiliary function calls (e.g., map.getCenter(), layer.getOpacity()), the system will automatically append the auxiliary function logic after executing the main function call. The output from these auxiliary functions will then be stored as part of the evaluation.

### 4.2 Judging Program

The overall process of the judging module is shown in **Figure 2c**, consisting of three stages: result reading, output parsing, and equivalence determination. The core task is to read the result file (specified by output_path) and, based on the output_type, select the appropriate comparison logic to evaluate the consistency between the model output and the expected_answer. Specifically, the 25 types covered by GeoJSEval-Bench encompass a broad range of types, from basic data structures to spatial geometric objects (as detailed in **Table 3**). However, in practice, although the configuration module specifies a clear output type for each function, results often exhibit overlap in both value and structural levels. For instance, Boolean and String both appear as atomic strings, while functions like geolib.coordinates and geolib.center output nested JSON objects, with differing field structures but similar semantics. Similarly, several spatial functions, such as turf.buffer, jsts.union, and turf.difference, return values in different forms, but essentially represent GeoJSON-wrapped spatial geometries, with types like Geometry, Feature, or FeatureCollection. Based on this, GeoJSEval introduces an intermediate abstraction layer for "numeric representations", merging types into several comparable data expression categories according to the actual format of the runtime results. The system then selects the most suitable comparison strategy based on this categorization. Finally, the system outlines the structural features and evaluation logic mapping for various function outputs across the five main libraries (as shown in **Table 4**). Based on this mapping, the framework establishes a type-aware equivalence judgment function, which selects the most appropriate comparison strategy based on the mapped type:

$$E(y,a) = \begin{cases} |y-a| < \epsilon, & if\ V = Number \\ y = a, & if\ V = String\ or\ Boolean \\ EqualSet(y,a), & if\ V = Array \\ StructMatch(y,a), & if\ V = JSON\ Object \\ GeoJSONMatch(y,a), & if\ V = GeoJSON \\ Eval(f(y)) = a, & if\ V = Map\ or\ Layer \end{cases} \tag{7}$$

Let $y$ denote the execution result of the model-generated code in the runtime environment, and $a$ represent the reference standard answer. $V$ represents the unified mapped performance type, and $\epsilon$ denotes the error tolerance threshold, primarily used for approximate matching of floating-point numeric types. "EqualSet$(y, a)$" is used for equivalence determination of set types, where the element order is ignored. "StructMatch$(y, a)$" is applicable to nested JSON structures, assessing key set consistency and exact key-value pair matching. "GeoJSONMatch$(y, a)$" is intended for standard GeoJSON spatial objects, evaluating the consistency of their geometric topology and attribute semantics. For complex object types returned by visualization platforms, such as Map or Layer, GeoJSEval cannot directly compare their entire structure. Instead, it introduces an auxiliary function set "$f(\cdot)$", such as "$f$ = {map.getCenter(), layer.getVisible()}", to extract key structural and semantic features from the model's execution results. The formal output "Eval$(f(y))$" is then compared with the reference answer $a$ for indirect equivalence determination. For visualization functions such as Leaflet and OpenLayers, GeoJSEval introduces an additional visualization verification mechanism: After the model code completes the webpage rendering, the system automatically captures the visible region of the page and saves it as a standard screenshot. Although screenshot results do not directly participate in scoring, they provide an auxiliary review perspective when the output structure is unclear or when the model's behavior may lead to visual discrepancies. This helps identify issues such as layers not loading, abnormal styling, or interaction failures—problems that structural analysis alone may fail to capture.

**Table 4.** Summary of Value Representations and Evaluation Strategies in GeoJSEval

| Source Library | Data Type | Value Representation | Testing Logic |
|---|---|---|---|
| Standard JavaScript types | Array | Array | Extract each value from the array and compare it one by one with the expected_answer. |
| Standard JavaScript types | Boolean | String | Compare each character with the answer one by one. |
| Standard JavaScript types | String | String | Compare each character with the answer one by one. |
| Standard JavaScript types | Number | Floating-point number | Directly compare with the answer. |
| Standard geometry types | Geometry | GeoJSON | For geometries, call Python's compare geometric consistency with Shapely; for features, extract geometry before comparison, then compare with the answer. |
| Standard geometry types | GeometryCollection | GeoJSON | For geometries, call Python's compare geometric consistency with Shapely; for features, extract geometry before comparison, then compare with the answer. |
| Standard geometry types | Feature | GeoJSON | For geometries, call Python's compare geometric consistency with Shapely; for features, extract geometry before comparison, then compare with the answer. |
| Standard geometry types | FeatureCollection | GeoJSON | For geometries, call Python's compare geometric consistency with Shapely; for features, extract geometry before comparison, then compare with the answer. |
| Geolib-specific | geolib.coordinates | JSON | Extract the corresponding key-value pairs and compare them one by one with the answer. |
| Geolib-specific | geolib.distanceCoordinate | JSON | Extract the corresponding key-value pairs and compare them one by one with the answer. |
| Geolib-specific | geolib.bounds | JSON | Extract the corresponding key-value pairs and compare them one by one with the answer. |
| Geolib-specific | geolib.center | JSON | Extract the corresponding key-value pairs and compare them one by one with the answer. |
| Leaflet-specific | leaflet.Map | JSON_indirect | Define the built-in methods of the Leaflet library for each function, convert them into a directly comparable JSON format, and compare with the answer. |
| Leaflet-specific | leaflet.Layer | JSON_indirect | Define the built-in methods of the Leaflet library for each function, convert them into a directly comparable JSON format, and compare with the answer. |
| Leaflet-specific | leaflet.Latlng | JSON | Extract the corresponding key-value pairs and compare them one by one with the answer. |
| Leaflet-specific | leaflet.LatlngBounds | JSON | Extract the corresponding key-value pairs and compare them one by one with the answer. |
| Leaflet-specific | leaflet.Point | JSON | Extract the corresponding key-value pairs and compare them one by one with the answer. |
| Leaflet-specific | leaflet.Bounds | JSON | Extract the corresponding key-value pairs and compare them one by one with the answer. |
| Openlayers-specific | ol.Map | JSON_indirect | Define the built-in methods of the Openlayers library for each function, convert them into a directly comparable JSON format, and compare with the answer. |
| Openlayers-specific | ol.View | JSON_indirect | Define the built-in methods of the Openlayers library for each function, convert them into a directly comparable JSON format, and compare with the answer. |
| Openlayers-specific | ol.Layer | JSON_indirect | Define the built-in methods of the Openlayers library for each function, convert them into a directly comparable JSON format, and compare with the answer. |
| Openlayers-specific | ol.Source | JSON_indirect | Define the built-in methods of the Openlayers library for each function, convert them into a directly comparable JSON format, and compare with the answer. |
| Openlayers-specific | ol.Coordinate | JSON | Extract the corresponding key-value pairs and compare them one by one with the answer. |
| Openlayers-specific | ol.Extent | JSON | Extract the corresponding key-value pairs and compare them one by one with the answer. |
| Openlayers-specific | ol.Size | JSON | Extract the corresponding key-value pairs and compare them one by one with the answer. |

## 5 Experiments

This chapter introduces the selection criteria for the models under evaluation, the experimental setup, and the evaluation metrics framework.

### 5.1 Evaluation Models

The evaluation objects selected in this study are mainstream LLMs publicly released as of July 2025, encompassing representative models with high performance, good reproducibility, and broad application impact. All included models are either officially open-source or have open interfaces, with stable release records and empirical validation in academic or industrial settings, ensuring the reproducibility and practical relevance of the evaluation results. The evaluation focuses on the model's performance in end-to-end application scenarios, aiming to provide researchers and developers in geospatial computing with practical guidance for model selection. It is important to note that although strategies like Prompt Engineering, RAG, and multi-agent collaboration may improve performance in some tasks, they do not fundamentally alter the model structure. Furthermore, their performance is highly dependent on specific configurations and scenario adaptation, raising concerns about stability and generalization capabilities. Additionally, such methods often introduce lengthy prompts, resulting in extra token consumption, which impacts the fairness and resource controllability of the evaluation. To ensure generalizability and cross-model comparability, these strategies are excluded from the current evaluation.

In terms of model type coverage, the evaluation encompasses four major categories: (1) general non-reasoning large language models; (2) general reasoning large language models; (3) general code generation models; (4) geospatial code generation models. For some models with publicly available versions of varying parameter sizes, this study tests different instances to comprehensively present the impact of model size on performance. A total of 18 model instances were evaluated, with the specific list of models and their parameter settings provided in **Table 5**

**Table 5.** Information of Evaluated LLMs

| Model Type | Model Name | Developer | Size | Year |
|---|---|---|---|---|
| General Non-Reasoning | DeepSeek-V3 | DeepSeek | 671B | 2024 |
| | GPT-4.1 | OpenAI | N/A | 2025 |
| | GPT-4.1-mini | OpenAI | N/A | 2025 |
| | Qwen-2.5 | Alibaba | 3B, 7B, 32B | 2024 |
| | Qwen-3 | Alibaba | 4B, 8B, 32B | 2025 |
| General Reasoning | o4-mini | OpenAI | N/A | 2025 |
| | QwQ-32B | Alibaba | 32B | 2025 |
| | DeepSeek-R1 | DeepSeek | 671B | 2025 |
| General Code Generation | DeepSeek-Coder-V2:16B | DeepSeek | 16B | 2024 |
| | Qwen2.5-Coder | Alibaba | 3B, 7B, 32B | 2024 |
| | Code-Llama-7B | Meta | 7B | 2023 |
| Geospatial Code Generation | GeoCode-GPT-7B | Wuhan University | 7B | 2024 |

### 5.2 Experimental Setup

The experiment was conducted on a local machine equipped with 32GB of memory and an NVIDIA RTX 4090 GPU, enabling efficient inference and concurrent processing for medium-scale models. For open-source models with parameter sizes not exceeding 16B, the Ollama tool was used to locally deploy and execute inference. For closed-source models with parameters exceeding 16B or those that do not support local deployment, inference was performed via their official API interfaces to ensure operational stability and output consistency. In terms of parameter settings, general-purpose non-reasoning models were configured with a low generation temperature (temperature = 0.2) to enhance output determinism. Large models with reasoning capabilities retained their default configurations to fully leverage their native logical abilities. To prevent output truncation and standardize result formats, the maximum output token count for all models was set to 16,384. The time consumption for each phase and task descriptions are detailed in **Table 6**.

**Table 6. Time Allocation Across Experimental Stages**

| Stages | Time Spent (hours) |
|---|---|
| GeoJSEval -Bench Construction | 45 |
| Expert Manual Revision | 65 |
| Model Inference and Code Execution | 360 |
| Evaluation of Model Responses | 290 |
| **Total (All Stages)** | **760** |

### 5.3 Evaluation Metrics

This study systematically evaluates the performance of LLMs in geospatial code generation tasks across four key dimensions: accuracy metrics, resource consumption metrics, operational efficiency metrics, and error type logs.

#### 5.3.1 Accuracy Metrics

To comprehensively evaluate the stability and accuracy of the model in code generation tasks, this study uses pass@n as the core performance metric. This metric indicates the probability that at least one of the outputs, after n independent generations for the same problem, matches the reference answer. It is commonly used to assess the reliability and robustness of models in generation tasks. Given that large language models often experience hallucination issues during generation—where multiple outputs with significant semantic or syntactic differences arise from the same input—single-round generation is often insufficient to reflect the true capabilities of the model. To enhance the stability of the evaluation and the reliability of the results, this study sets three configurations for comparison, with n = 1, 3, and 5.

$$pass@n = 1 - \frac{\binom{N-c}{n}}{\binom{N}{n}} \tag{8}$$

Here, $N$ represents the total number of generated samples, and $c$ denotes the number of correct results among them.

Furthermore, this study introduces the Coefficient of Variation (CV) as an additional evaluation metric to measure the degree of variation in the model's results across multiple generations, indirectly reflecting the severity of the hallucination phenomenon. The CV is defined as the ratio of the standard deviation to the mean, with the following formula:

$$CV = \frac{\sigma}{\mu} \tag{9}$$

Here, $\sigma$ is the standard deviation, and $\mu$ is the mean. A lower CV value indicates less variability in the model's generated results, reflecting higher stability.

To achieve a more comprehensive model evaluation, this study further introduces Stability-Adjusted Accuracy (SA), which builds on the evaluation of final accuracy (represented by Pass@5) while incorporating the stability factor (CV) into the calculation. A higher SA value is achieved when Pass@5

scores are high and CV is low, reflecting that the model not only maintains accuracy but also exhibits strong consistency in generation. The calculation is as follows:

$$SA = \frac{Pass@5}{1 + CV} \tag{10}$$

### 5.3.2 Resource Consumption Metrics

The resource consumption metrics aim to assess the utilization of computational power and resources by LLMs during geospatial code generation tasks. These metrics include three dimensions:

Token Consumption (Tok.): Represents the average number of tokens consumed by the model to complete each unit test. For locally deployed open-source models, this value reflects their usage of local computational resources, such as memory and GPU memory. When calling closed-source or commercial model APIs, the token count is directly mapped to usage costs. Currently, most mainstream models charge based on "tokens per million", with significant variations in consumption. This impacts resource scheduling, as well as the economic feasibility and scalability of the system.

Inference Time (In.T): Refers to the average response time (in seconds) for the model to generate the code for a single test case. It is used to evaluate the model's inference latency and response efficiency, which directly impacts the user interaction experience in real-world applications.

Code Lines (Co.L): This metric counts the number of valid code lines in the model's output, excluding non-functional content such as comments, natural language explanations, and formatting instructions. Compared to token consumption, this metric better reflects the model's actual output in structured code generation, making it easier to assess its programming efficiency and output quality.

### 5.3.3 Operational Efficiency Metrics

To systematically evaluate the cost-effectiveness of large language models in geospatial code generation tasks, this study introduces the operational efficiency metrics to measure the level of accuracy achieved per unit of resource consumption. These metrics are based on three key resource dimensions: time consumption, token consumption, and code structure complexity, which correspond to the model's inference efficiency, token utilization efficiency, and structured code efficiency, respectively. Considering the inherent randomness in model outputs, and that each task generates five results, all efficiency-related metrics are based on Pass@5, ensuring comparability and statistical robustness across models in terms of evaluation frequency and accuracy.

Inference Efficiency: Inference efficiency measures the average accuracy achieved by the model per unit of time. It is calculated as the accuracy divided by the average inference time (in seconds):

$$\text{Inference Efficiency} = \frac{\text{pass}@n}{\text{Inference Time}} \tag{11}$$

This metric reflects the model's overall balance between inference speed and output quality. A shorter inference time combined with higher accuracy indicates a stronger advantage in computational resource utilization and user response experience, making it suitable for tasks that require high real-time interactivity.

Token Efficiency: Token efficiency characterizes the accuracy achieved by the model per unit of token cost,

defined as follows:

$$\text{Token Efficiency} = \frac{\text{pass@}n}{\text{Token Consumption}} \tag{12}$$

This metric can be used to compare the cost-effectiveness between different models, particularly useful for evaluating API-based pay-per-use large language models. It holds significant reference value in scenarios with limited resource budgets or large-scale deployments. Models with lower token consumption and higher accuracy exhibit a better cost-performance ratio (cost-effectiveness).

Code Line Efficiency: Code efficiency focuses on the number of core executable code lines generated by the model, excluding non-structural content such as natural language explanations and comments. It is defined by the following formula:

$$\text{Code Efficiency} = \frac{\text{pass@}n}{\text{CodeLines}} \tag{13}$$

This metric reveals the model's ability to generate output with compact structure and logical validity, closely aligning with the requirements for maintainability and execution efficiency in real-world engineering development. Models that generate concise and structurally clear code have greater practical value during deployment and subsequent iterations.

### 5.3.4 Error type logs

To further enhance the qualitative understanding of model behavior and the identification of performance bottlenecks, the GeoJSEval framework introduces an automatic runtime error capture mechanism in the standard JavaScript execution environment. This mechanism systematically records the types of errors generated during the code generation process by large language models, aiding in targeted optimization of model performance. This mechanism supports the identification and classification of the following four common error types:

Syntax Error: Refers to issues in the syntactical structure of the code that prevent it from compiling. For example, unclosed parentheses, spelling mistakes, or missing required module imports, which commonly result in a SyntaxError.

Attribute or Parameter Error: Refers to incorrect calls to object properties or function parameters, such as accessing a non-existent function or passing incorrect types or quantities of parameters. This typically results in a TypeError or ReferenceError. This type of error indicates a misunderstanding by the model of the API semantics and interface structure, especially in handling function signatures and invocation constraints.

Output Type Error: Refers to cases where the result type returned by the generated code after execution does not match the expected type and cannot be correctly parsed by the evaluation program. For example, expecting a GeoJSON.Feature but the actual output is a general object, or mistakenly generating a string instead of a coordinate array. This type of error reflects the model's shortcomings in type inference, structure construction, and task semantic understanding, directly affecting the code's evaluability and generality.

Invalid Answer: Refers to cases where the code syntax and types are correct, and the code executes successfully, but the output does not match the reference answer. This may manifest as logical errors in functionality implementation or incorrect function selection. This type of error is attributed to the model's misunderstanding of the task objectives or function semantics, representing a semantic-level failure.

Runtime Error: Refers to cases where the model-generated code experiences prolonged periods of no response or enters an infinite loop during execution, exceeding the system's maximum runtime threshold (30 seconds), resulting in a forced termination. These errors are often caused by irrational model logic design, unterminated asynchronous function calls, repeated recursion, or excessive data processing. They severely impact the executability of the task and the stability of system resource utilization.

Other Error: This type of error refers to miscellaneous issues that do not fall under common syntax, parameter, or runtime exceptions. These errors typically arise from the model's misinterpretation or neglect of contextual prompts during the code generation process. For example, in the GeoJSEval testing environment, all necessary third-party libraries (such as Turf, Geolib, OpenLayers, Leaflet, etc.) are pre-loaded, and the prompt explicitly informs the model that "module imports do not need to be repeated." Despite this, some models still exhibit the behavior of redundantly importing modules when generating code, such as re-invoking import or require statements. Although this type of error does not affect the syntax itself, it can lead to module conflicts or duplicate declaration errors in the JavaScript execution environment, affecting code executability and evaluation stability. This phenomenon reflects a gap in current large language models' ability to understand multi-turn contexts and follow instructions accurately.

## 6. Results

Building upon Section 5.3, this chapter systematically introduces and analyzes the evaluation results based on the GeoJSEval framework and the GeoJSEval-Bench test suite, focusing on four key aspects: accuracy metrics, resource consumption metrics, operational efficiency metrics, and error type logs.

### 6.1 Accuracy

**Figure 8** shows the performance of various large language models in terms of execution accuracy across five major JavaScript geospatial libraries, with pass@5 as the primary metric. Overall, there are significant differences in the average accuracy rates of models across different libraries, reflecting an uneven generalization ability across task types. Models performed better in libraries focused on visualization and interaction control, such as Leaflet (0.837) and OpenLayers (0.633), indicating that they are more adaptable to tasks with intuitive semantics and structured frameworks. However, in libraries like Turf.js (0.571) and Geolib (0.565), which involve complex spatial logic and geometric computations, the accuracy is relatively low, indicating that the models still have significant shortcomings in spatial semantic modeling. From a model category perspective, General Reasoning models consistently maintain high accuracy across all libraries, demonstrating stronger cross-task stability and structural adaptability. In contrast, General Non-Reasoning and Geospatial Code Generation models show a noticeable drop in performance on tasks with complex structures or ambiguous function semantics. Individual models, such as Qwen3-32B, show considerable variation in accuracy across different libraries, indicating strong task sensitivity. GeoCode-GPT and CodeLlama-7B rank lowest in accuracy across all libraries, indicating that they lack effective generalization capabilities for the semantic structures of mainstream JS geospatial libraries.

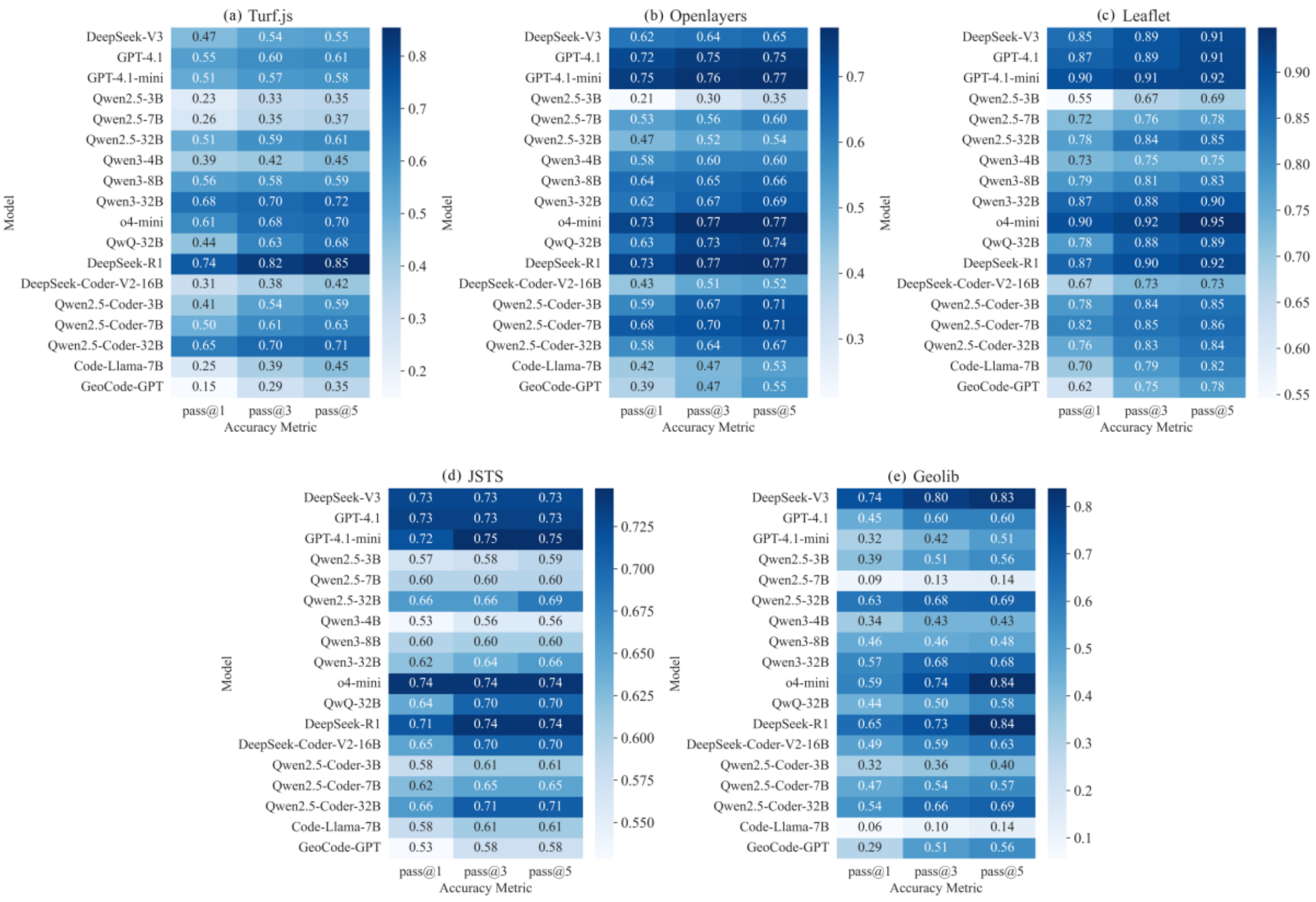

**Figure 8. Heatmap of Accuracy Across Models and Geospatial JavaScript Libraries.** Each cell represents the pass@1, pass@3, pass@5 accuracy (%) of a specific model on a specific geospatial JavaScript library. Darker shades indicate higher accuracy. The white text within each cell displays the exact pass@1, pass@3, pass@5 score.

The accuracy-related evaluation results for each model are shown in **Table 7**. The overall pass@n metric for each model is calculated by combining all tasks across the five geospatial libraries, rather than averaging the results for each library, to more accurately reflect the model's overall performance in JavaScript geospatial code understanding and generation tasks.

**Table 7. Accuracy Evaluation Results,** where the values in parentheses under Pass@3 represent the improvement over Pass@1, and the values in parentheses under Pass@5 represent the improvement over Pass@3.

| Model | pass@1(%) | pass@3(%) | pass@5(%) | CV | SA |
|---|---|---|---|---|---|
| **General Non-Reasoning** | | | | | |
| DeepSeek-V3 | 65.69 | 69.72 (+4.03) | 71.16 (+1.44) | 0.034 | 68.85 |
| GPT-4.1 | 69.05 | 72.84 (+3.79) | 73.90 (+1.06) | 0.029 | 71.82 |
| GPT-4.1-mini | 67.99 | 71.69 (+3.70) | 72.89 (+1.20) | 0.029 | 70.81 |
| Qwen2.5-3B | 35.75 | 45.35 (+9.60) | 48.03 (+2.68) | 0.122 | 42.79 |
| Qwen2.5-7B | 47.17 | 52.30 (+5.13) | 54.22 (+1.92) | 0.058 | 51.24 |
| Qwen2.5-32B | 59.40 | 65.36 (+5.96) | 67.18 (+1.82) | 0.052 | 63.86 |
| Qwen3-4B | 53.79 | 56.86 (+3.07) | 58.06 (+1.20) | 0.032 | 56.26 |
| Qwen3-8B | 63.63 | 65.45 (+1.82) | 66.70 (+1.25) | 0.019 | 65.44 |
| Qwen3-32B | 70.11 | 73.32 (+3.21) | 75.14 (+1.82) | 0.029 | 73.05 |
| **General Reasoning** | | | | | |
| o4-mini | 72.84 | 77.93 (+5.09) | 80.04 (+2.11) | 0.039 | 77.01 |
| QwQ-32B | 59.88 | 71.74 (+11.86) | 74.57 (+2.83) | 0.093 | 68.25 |
| DeepSeek-R1 | 76.20 | 81.43 (+5.23) | 83.78 (+2.35) | 0.039 | 80.61 |
| **General Code Generation** | | | | | |
| DeepSeek-Coder-V2-16B | 48.66 | 55.33 (+6.67) | 57.58 (+2.25) | 0.070 | 53.80 |
| Qwen2.5-Coder-3B | 56.53 | 64.88 (+8.35) | 67.56 (+2.68) | 0.075 | 62.87 |
| Qwen2.5-Coder-7B | 64.06 | 69.34 (+5.28) | 70.97 (+1.63) | 0.043 | 68.02 |
| Qwen2.5-Coder-32B | 65.21 | 71.69 (+6.48) | 73.32 (+1.63) | 0.050 | 69.83 |
| Code-Llama-7B | 43.19 | 51.49 (+8.30) | 56.14 (+4.65) | 0.107 | 50.73 |
| **Geospatial Code Generation** | | | | | |
| GeoCode-GPT | 38.53 | 50.48 (+11.95) | 55.42 (+4.94) | 0.147 | 48.31 |

**Figure 9** shows the performance differences of 18 models across pass@1, pass@3, and pass@5, reflecting their accuracy distribution characteristics in both single-round and multi-round generation processes. Overall, DeepSeek-R1 (83.8%), o4-mini (80.0%), and Qwen3-32B (75.1%) performed the best. DeepSeek-R1 achieved pass@1 of 76.2%, with limited subsequent gains, indicating strong one-shot generation capability. Model size is closely related to performance. In the Qwen-2.5 and Qwen-2.5-Coder series, the 32B model outperforms the 7B and 3B versions significantly, suggesting that parameter scaling improves stability. It is worth noting that models such as GPT-4.1, GPT-4.1-mini, and DeepSeek-V3 achieved high accuracy at pass@1, reflecting their good generalization ability and high initial generation quality in geospatial tasks.

**Figure 10** analyzes the accuracy improvement in multi-round generation, reflecting the "compensability" of candidate redundancy. The results show that the average gain from pass@1 to pass@3 is 6.1%, while the gain from pass@3 to pass@5 is only 2.2%, displaying a clear diminishing marginal effect. This suggests that most effective candidates are concentrated in the first three generations. Some models, such as GeoCode-GPT (+12.0%), Qwen3-32B (+11.9%), and Qwen-2.5-3B (+9.6%), show significant improvement at pass@3, indicating that their initial generation is unstable but has the potential to achieve better solutions through candidate reordering. In contrast, models like the GPT-4.1 series and DeepSeek-V3 show smaller gains, indicating that their initial generation quality is high and the value of candidate redundancy is limited. Multi-round generation strategies have a stronger compensatory effect on smaller and medium-sized models, while the potential for improvement is more limited for larger models with higher structural optimization.

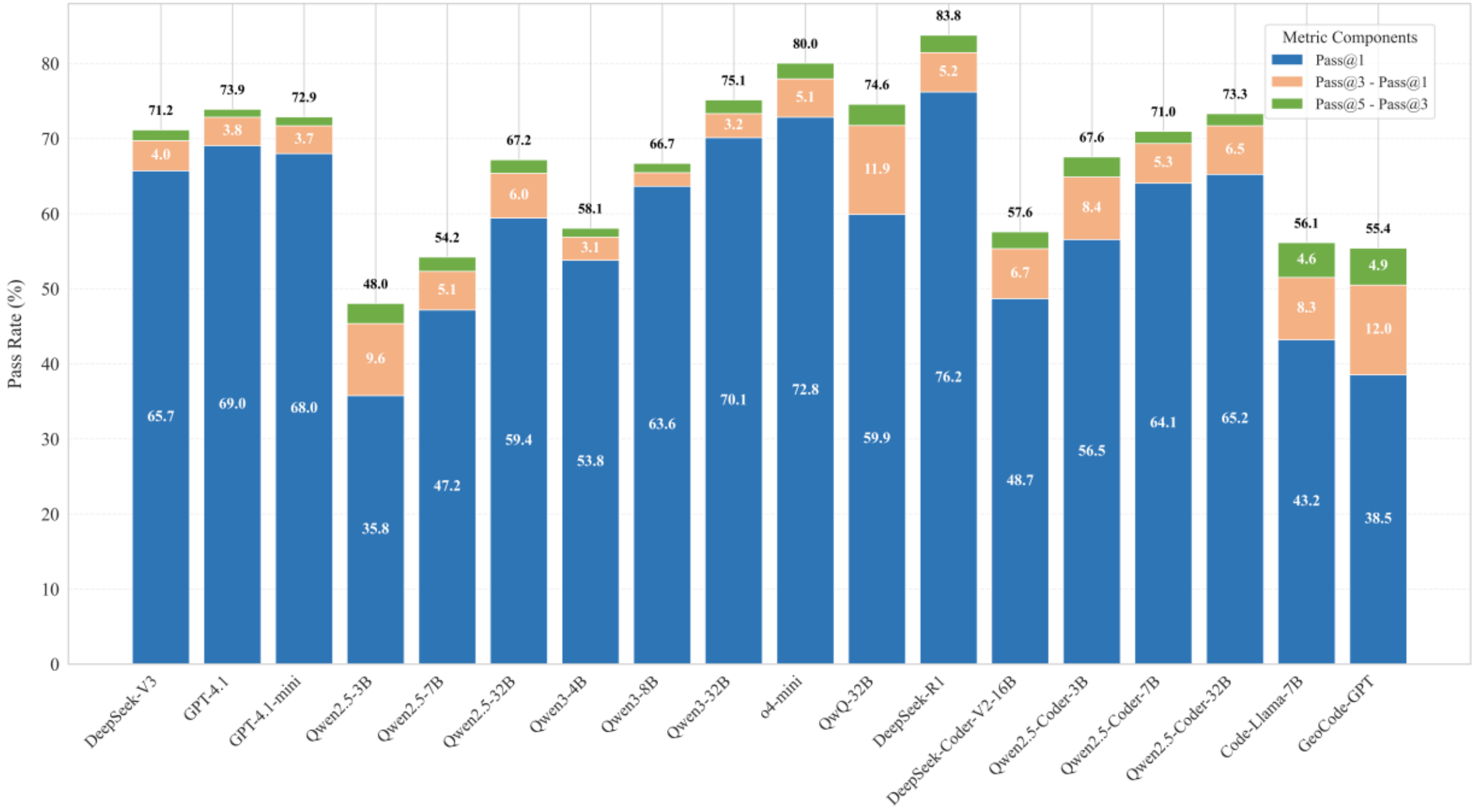

**Figure 9. Stacked Bar Chart of pass@n Metrics.** The blue represents the Pass@1 value, the orange represents the improvement of Pass@3 over Pass@1, and the green represents the improvement of Pass@5 over Pass@3. The white text on the bars indicates the absolute scores for Pass@1, Pass@3, and Pass@5, respectively.

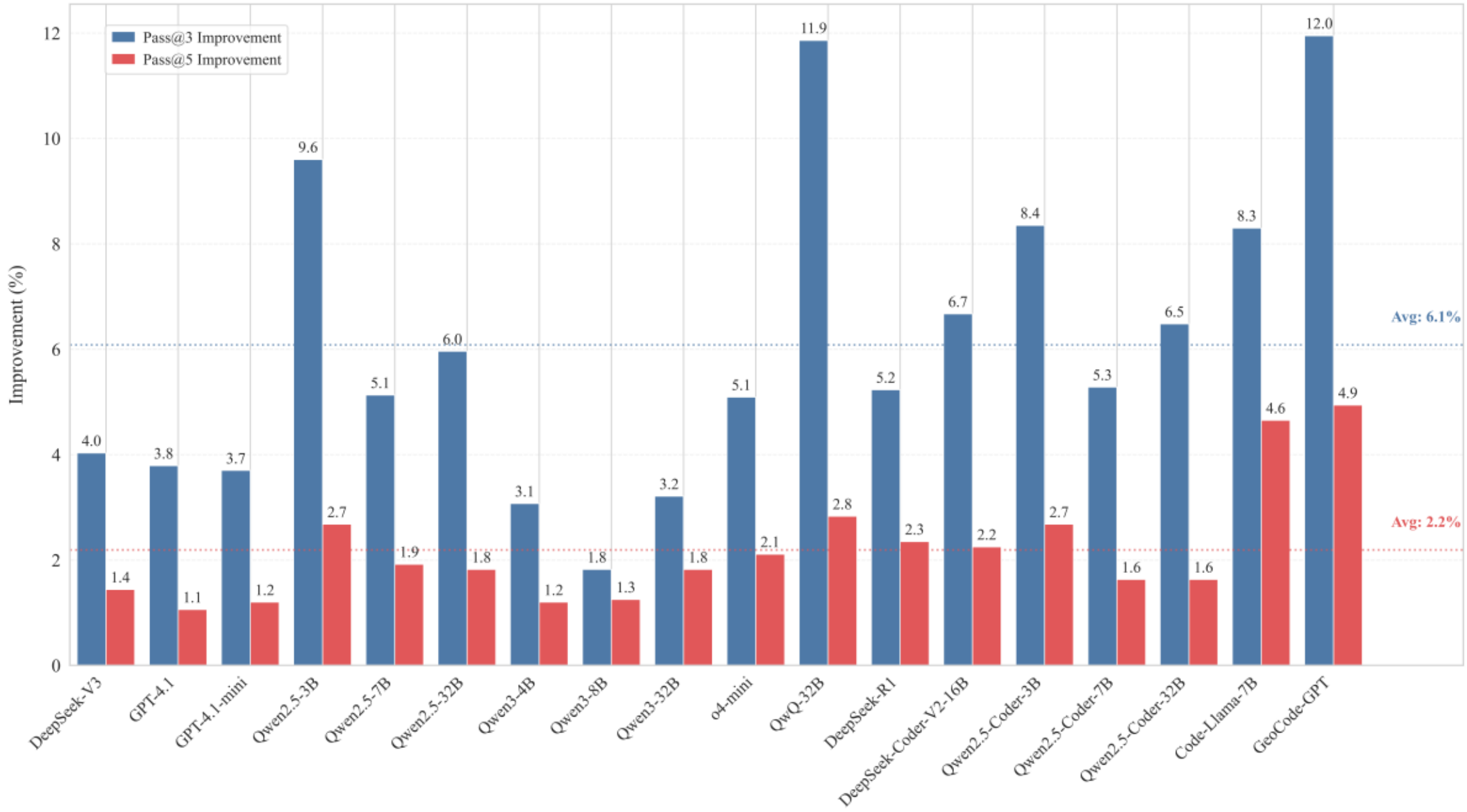

**Figure 10. Bar Chart of Pass@3 and Pass@5 Improvement Ratios.** The blue and red bars represent the improvement of Pass@3 over Pass@1 and Pass@5 over Pass@3, respectively. Dotted lines indicate the average gains across all models.

To evaluate the trade-off between accuracy and stability in code generation by large language models, this study constructs a composite ranking metric based on Pass@5, Coefficient of Variation (CV), and Stability-Adjusted Accuracy (SA), with the results shown in **Table 8**. The orange shading in the table indicates models with higher accuracy ranking (P_Rank) but poorer stability, leading to a lower composite ranking (S_Rank). Examples include QwQ-32B, Qwen2.5-Coder-3B, and Code-Llama-7B, reflecting their significant fluctuations in generated results. The blue shading indicates models with average accuracy rankings but outstanding stability, showing a clear advantage in S_Rank. Examples include GPT-4.1, GPT-4.1-mini, DeepSeek-V3, Qwen2.5-7B, and Qwen3-8B, making them suitable for scenarios where consistency in results is crucial. General Reasoning models demonstrate a more balanced performance between accuracy and stability, leading to higher composite rankings.

**Table 8. Ranking of the models under Pass@5, CV, and SA metrics.** P_Rank, C_Rank, and S_Rank represent the rankings based on Pass@5, CV, and SA, respectively. Higher values of Pass@5 and SA indicate better performance and higher ranking, while a lower CV value indicates better performance and a higher ranking. The table is sorted by S_Rank, reflecting the accuracy ranking of the models with the inclusion of stability factors, rather than solely considering accuracy. Category 1, 2, 3, and 4 correspond to General Non-Reasoning Models, General Reasoning Models, General Code Generation Models, and Geospatial Code Generation Models, respectively.

| Category | Model | pass@5 | CV | SA | P_Rank | C_Rank | S_rank |
|---|---|---|---|---|---|---|---|
| 2 | Deepseek-R1 | 0.8378 | 0.039 | 0.81 | 1 | 8 | 1 |
| 2 | o4-mini | 0.8004 | 0.039 | 0.77 | 2 | 7 | 2 |
| 1 | Qwen3-32B | 0.7514 | 0.029 | 0.73 | 3 | 2 | 3 |
| 1 | GPT-4.1 | 0.739 | 0.029 | 0.72 | 5 | 3 | 4 |
| 1 | GPT-4.1-mini | 0.7289 | 0.029 | 0.71 | 7 | 4 | 5 |
| 3 | Qwen2.5-Coder-32B | 0.7332 | 0.050 | 0.70 | 6 | 10 | 6 |
| 1 | Deepseek-V3 | 0.7116 | 0.034 | 0.69 | 8 | 6 | 7 |
| 2 | QwQ-32B | 0.7457 | 0.093 | 0.68 | 4 | 15 | 8 |
| 3 | Qwen2.5-Coder-7B | 0.7097 | 0.043 | 0.68 | 9 | 9 | 9 |
| 1 | Qwen3-8B | 0.667 | 0.019 | 0.65 | 12 | 1 | 10 |
| 1 | Qwen2.5-32B | 0.6718 | 0.052 | 0.64 | 11 | 11 | 11 |
| 3 | Qwen2.5-Coder-3B | 0.6756 | 0.075 | 0.63 | 10 | 14 | 12 |
| 1 | Qwen3-4B | 0.5806 | 0.032 | 0.56 | 13 | 5 | 13 |
| 3 | Deepseek-Coder-V2-16B | 0.5758 | 0.070 | 0.54 | 14 | 13 | 14 |
| 1 | Qwen2.5-7B | 0.5422 | 0.058 | 0.51 | 17 | 12 | 15 |
| 3 | Code-Llama-7B | 0.5614 | 0.107 | 0.51 | 15 | 16 | 16 |
| 4 | Geocode-GPT | 0.5542 | 0.147 | 0.48 | 16 | 18 | 17 |
| 1 | Qwen2.5-3B | 0.4803 | 0.122 | 0.43 | 18 | 17 | 18 |

### 6.2 Resource Consumption

The evaluation results of resource consumption are shown in **Table 9**, where this study visualizes the analysis of token consumption, inference time, and the number of core code lines generated.

**Table 9. Evaluation Results for Resource Consumption.** For the QwQ-32B model using API calls, due to the provider's configuration, only "streaming calls" are supported. In this mode, Token consumption cannot be tracked, so it is marked as N/A.

| Model | Inference Method | Tok.(tokens) | In. T(s) | Co. L(lines) |
|---|---|---|---|---|
| **General Non-Reasoning** | | | | |
| Deepseek-V3 | API call | 484 | 10.90 | 17.56 |
| GPT-4.1 | API call | 631 | 4.28 | 32.80 |
| GPT-4.1-mini | API call | 701 | 5.26 | 35.90 |
| Qwen2.5-3B | Local deployment | 514 | 0.80 | 13.83 |
| Qwen2.5-7B | Local deployment | 421 | 0.51 | 7.22 |
| Qwen2.5-32B | Local deployment | 502 | 3.90 | 14.72 |
| Qwen3-4B | Local deployment | 489 | 0.75 | 11.70 |
| Qwen3-8B | Local deployment | 535 | 1.42 | 15.92 |
| Qwen3-32B | Local deployment | 521 | 4.38 | 14.52 |
| **General Reasoning** | | | | |
| o4-mini | API call | 1507 | 11.81 | 18.21 |
| QwQ-32B | Local deployment | N/A | 41.35 | 32.29 |
| Deepseek-R1 | API call | 2958 | 111.97 | 48.46 |
| **General Code Generation** | | | | |
| Deepseek-Coder-V2-16B | Local deployment | 544 | 0.87 | 13.82 |
| Qwen2.5-Coder-3B | Local deployment | 541 | 0.92 | 17.00 |
| Qwen2.5-Coder-7B | Local deployment | 421 | 0.51 | 7.29 |
| Qwen2.5-Coder-32B | Local deployment | 490 | 3.57 | 13.69 |
| Code-Llama-7B | Local deployment | 497 | 0.56 | 7.78 |
| **Geospatial Code Generation** | | | | |
| GeoCode-GPT | Local deployment | 643 | 2.21 | 28.14 |

**Figure 11** displays the average token consumption of each model during the code generation process, with the green line representing the average pass@5 performance across the five geospatial libraries. The results show that General Reasoning models consume far more tokens than other model types, approximately 15 to 42 times higher than General Non-Reasoning, Code Generation, and Geospatial Code Generation models. Despite their clear advantage in accuracy, the high inference costs increase the usage expenses. This result provides a cost assessment basis for model deployment, advising users to balance accuracy benefits with resource consumption when selecting models, and to match the application scenario with performance requirements accordingly.

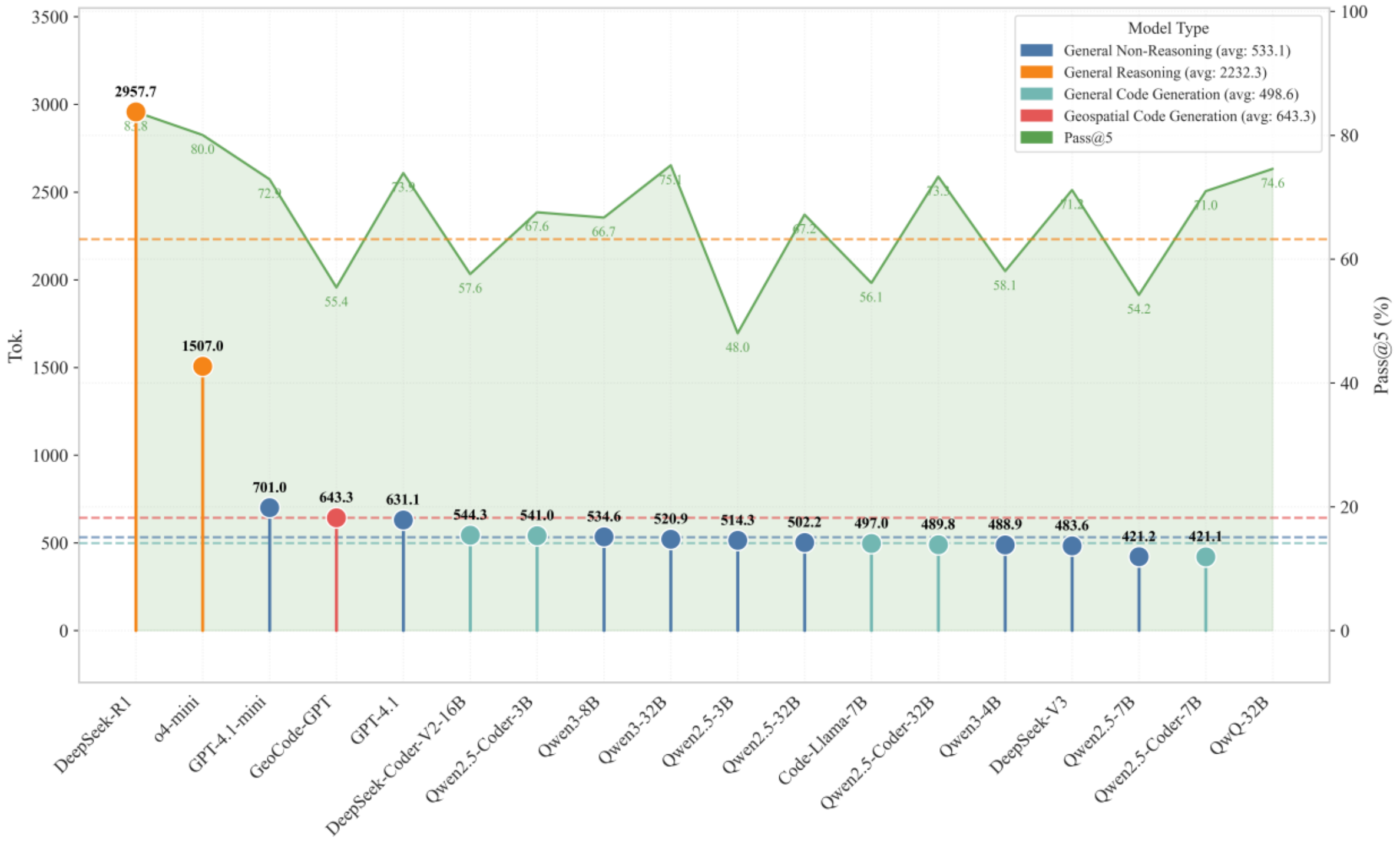

**Figure 11. Average Token Consumption of LLMs and Mean pass@5 Performance Across Five JavaScript Libraries for Different LLMs.** Each point on the green line represents the pass@5 value of the corresponding model.

The inference time consumption of various large language models during code generation is shown in **Figure 12**. General Reasoning models generally have higher inference times, but o4-mini is an exception, with an average inference time lower than most models, even slightly better than DeepSeek-V3, possibly benefiting from a more efficient inference architecture or deployment optimization. The average inference times of DeepSeek-R1 and QwQ-32B are 112.0 seconds and 41.3 seconds, respectively, significantly higher than similar models, indicating substantial inference overhead. This result suggests that in resource-constrained or response-sensitive application scenarios, the practicality of such models should be carefully evaluated, and inference efficiency can be further optimized in the future in line with deployment requirements.

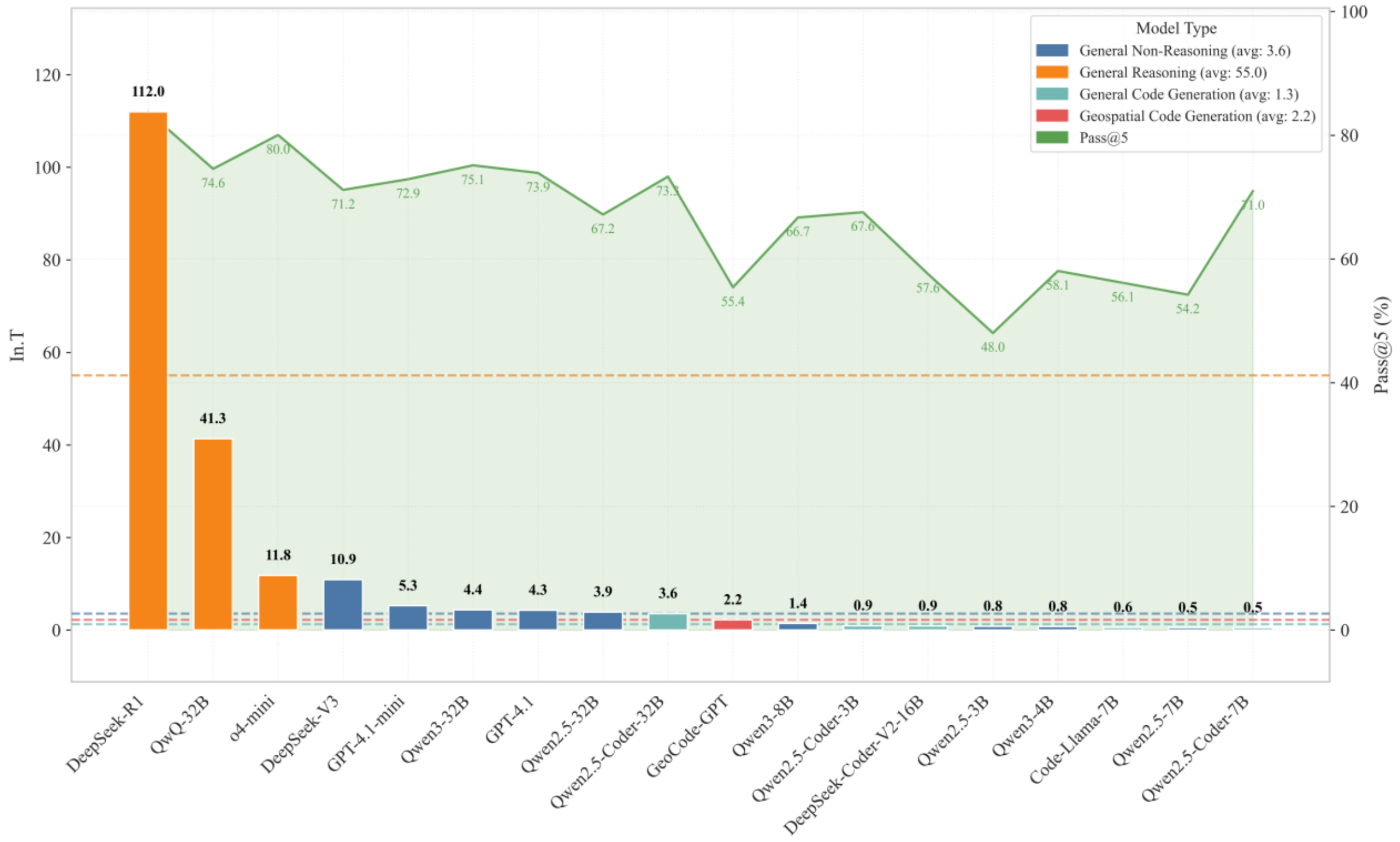

**Figure 12. Average Inference Time Comparison Acorss LLMs and Mean pass@5 Performance Across Five JavaScript Libraries for Different LLMs.**

To analyze the model output features, this study uses total lines of response (Res.L) and cleaned code lines (Co.L) as evaluation metrics, with the results shown in **Figure 13**. Res.L shows that General Reasoning models have higher overall redundancy, with DeepSeek-R1 and QwQ-32B being particularly notable, primarily due to their inclusion of the complete reasoning chain content. In contrast, o4-mini has a significantly lower Res.L as it does not explicitly output the reasoning process. The Co.L metric shows that DeepSeek-R1, the GPT-4.1 series, and QwQ-32B generate more code lines, possibly related to the embedding of comments and structural normalization. Notably, in the Qwen2.5 series, code length does not increase linearly with parameter size, and some larger models even generate more concise code, reflecting their stronger output compression and structural abstraction capabilities developed during training.

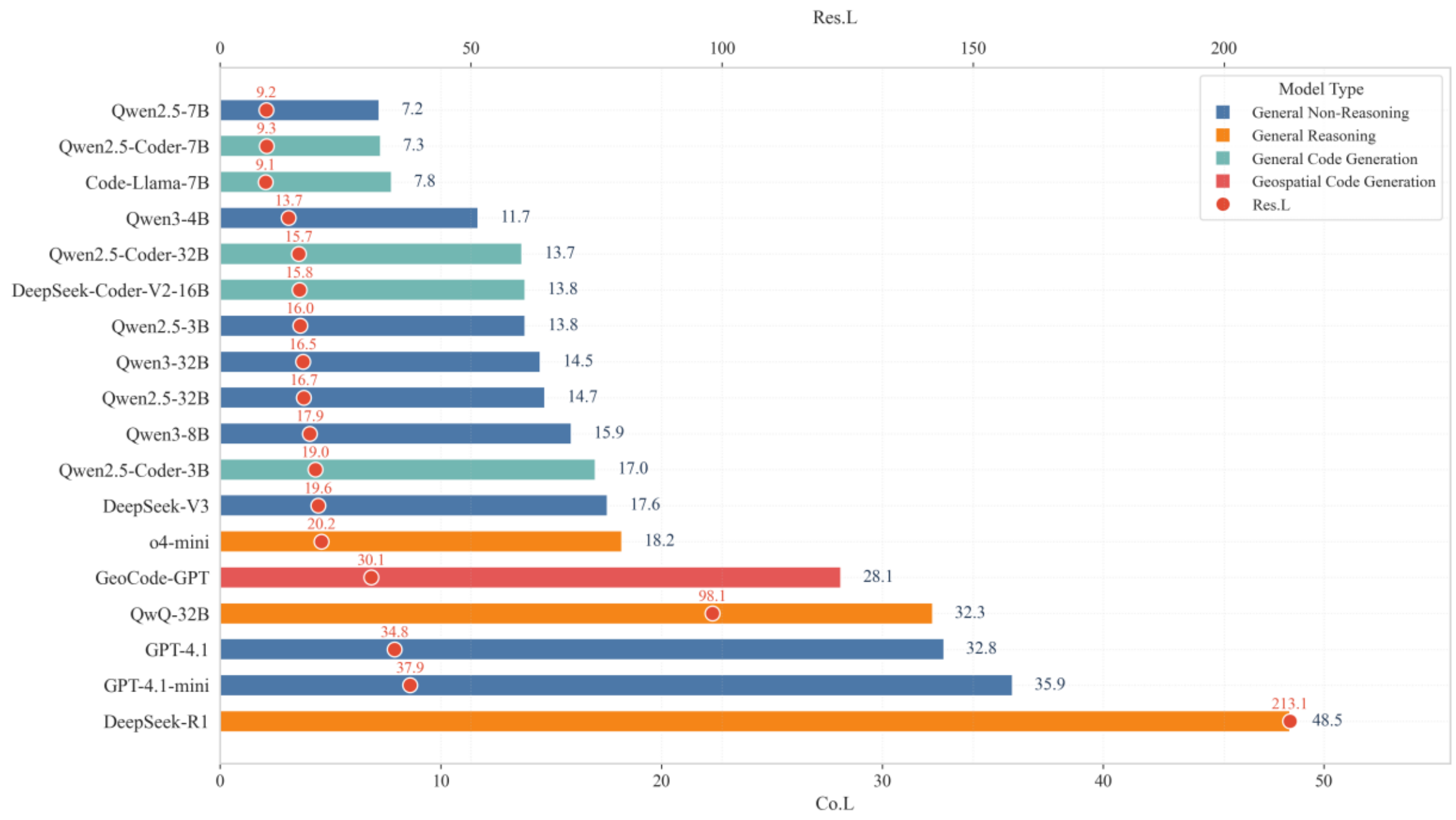

**Figure 13. Average Lines of Generated Code per Model.** The black numbers indicate the values of Co.L, while the red numbers represent the values of Res.L.

### 6.3 Operational Efficiency

The operational efficiency results for each model are shown in **Table 10**.

**Table 10. Evaluation Results for Operational efficiency.** For the QwQ-32B model using API calls, due to the provider's configuration, only "streaming calls" are supported. In this mode, Token consumption cannot be tracked, so it is marked as N/A.

| Model | Inference Method | Tok.-E | In.T-E | Co.L-E |
|---|---|---|---|---|
| **General Non-Reasoning** | | | | |
| Deepseek-V3 | API call | 0.0015 | 0.065 | 0.041 |
| GPT-4.1 | API call | 0.0012 | 0.172 | 0.023 |
| GPT-4.1-mini | API call | 0.0010 | 0.139 | 0.020 |
| Qwen2.5-3B | Local deployment | 0.0009 | 0.600 | 0.035 |
| Qwen2.5-7B | Local deployment | 0.0013 | 1.057 | 0.075 |
| Qwen2.5-32B | Local deployment | 0.0013 | 0.172 | 0.046 |
| Qwen3-4B | Local deployment | 0.0012 | 0.773 | 0.050 |
| Qwen3-8B | Local deployment | 0.0012 | 0.469 | 0.042 |
| Qwen3-32B | Local deployment | 0.0014 | 0.172 | 0.052 |
| **General Reasoning** | | | | |
| o4-mini | API call | 0.0005 | 0.068 | 0.044 |
| QwQ-32B | Local deployment | N/A | 0.018 | 0.023 |
| Deepseek-R1 | API call | 0.0003 | 0.007 | 0.017 |
| **General Code Generation** | | | | |
| Deepseek-Coder-V2-16B | Local deployment | 0.0011 | 0.663 | 0.042 |
| Qwen2.5-Coder-3B | Local deployment | 0.0012 | 0.731 | 0.040 |
| Qwen2.5-Coder-7B | Local deployment | 0.0017 | 1.387 | 0.097 |
| Qwen2.5-Coder-32B | Local deployment | 0.0015 | 0.205 | 0.054 |

| Model | Inference Method | Tok.-E | In.T-E | Co.L-E |
|---|---|---|---|---|
| Code-Llama-7B | Local deployment | 0.0011 | 0.999 | 0.072 |
| | **Geospatial Code Generation** | | | |
| GeoCode-GPT | Local deployment | 0.0009 | 0.251 | 0.020 |

Since Tok.-E, In.T-E, and Co.L-E are ratio-based efficiency metrics, lacking a unified dimension and intuitive interpretability, this study independently ranks them as T_Rank, I_Rank, and Co_Rank, and computes the arithmetic average of the three to construct the composite efficiency ranking metric E_Rank, which measures the model's overall performance in terms of resource consumption (see **Table 11**). Combining P_Rank (accuracy ranking) and S_Rank (stability-adjusted accuracy ranking), a slice analysis of the three core ranking metrics is performed, and Table 10 is constructed accordingly. This provides a systematized evaluation of the relative advantages of each model in terms of accuracy, stability, and efficiency. It should be noted that P_Rank reflects the original accuracy ranking, S_Rank incorporates stability adjustments, and E_Rank reflects the correction effect of resource utilization efficiency on availability. Finally, the arithmetic average of these three is used to define the overall performance metric, Total_Rank, which is used to assess the model's overall performance. According to the ranking results, Qwen3-32B and Qwen2.5-Coder-32B rank highly across all three dimensions, showing stable performance and supporting local deployment, making them highly practical. In contrast, o4-mini, DeepSeek-R1, and QwQ-32B, which are general reasoning models, excel in accuracy and stability but have relatively lower efficiency rankings. They are more suitable for application scenarios that require high generation quality and are tolerant of higher latency.

**Table 11. Rank-Based Comparative Evaluation of Models.** The table is sorted by Total_Rank in ascending order. If models share the same average rank, they are assigned the same ranking (e.g., both DeepSeek-V3 and Code-Llama-7B are ranked 1 in E_Rank). Blue highlights indicate the top 12 models in E_Rank, green for the top 12 in S_Rank, and orange for the top 12 in P_Rank. Gray highlights mark the bottom 6 models across E_Rank, S_Rank, and P_Rank. Categories 1, 2, 3, and 4 correspond to General Non-Reasoning Models, General Reasoning Models, General Code Generation Models, and Geospatial Code Generation Models, respectively.

| Category | Model | T_Rank | I_Rank | Co_Rank | E_Rank | S_rank | P_Rank | Total_Rank |
|---|---|---|---|---|---|---|---|---|
| 1 | Qwen3-32B | 4 | 13 | 5 | 6 | 3 | 3 | **1** |
| 3 | Qwen2.5-Coder-32B | 2 | 10 | 4 | 3 | 6 | 6 | **2** |
| 2 | o4-mini | 16 | 15 | 8 | 14 | 2 | 2 | **3** |
| 3 | Qwen2.5-Coder-7B | 1 | 1 | 1 | 1 | 9 | 9 | **4** |
| 2 | Deepseek-R1 | 17 | 18 | 18 | 18 | 1 | 1 | **5** |
| 1 | GPT-4.1 | 10 | 11 | 15 | 13 | 4 | 5 | **6** |
| 1 | Deepseek-V3 | 3 | 16 | 11 | 11 | 7 | 8 | **7** |
| 1 | GPT-4.1-mini | 13 | 14 | 16 | 16 | 5 | 7 | **8** |
| 1 | Qwen2.5-32B | 5 | 12 | 7 | 7 | 11 | 11 | **9** |
| 2 | QwQ-32B | 18 | 17 | 14 | 17 | 8 | 4 | **9** |
| 3 | Qwen2.5-Coder-3B | 7 | 5 | 12 | 7 | 12 | 10 | **9** |
| 1 | Qwen3-4B | 9 | 4 | 6 | 5 | 13 | 13 | **12** |
| 1 | Qwen3-8B | 8 | 8 | 9 | 9 | 10 | 12 | **12** |
| 1 | Qwen2.5-7B | 6 | 2 | 2 | 2 | 15 | 17 | **14** |
| 3 | Code-Llama-7B | 11 | 3 | 3 | 4 | 16 | 15 | **15** |

| Category | Model | T_Rank | I_Rank | Co_Rank | E_Rank | S_rank | P_Rank | Total_Rank |
|---|---|---|---|---|---|---|---|---|
| 3 | Deepseek-Coder-V2-16B | 12 | 6 | 10 | 10 | 14 | 14 | **16** |
| 1 | Qwen2.5-3B | 14 | 7 | 13 | 12 | 18 | 18 | **17** |
| 4 | GeoCode-GPT | 15 | 9 | 17 | 15 | 17 | 16 | **17** |

### 6.4 Error Type Analysis

The types of errors encountered by each model during code generation are shown in **Table 12**. It can be seen that runtime errors and other errors are sporadically distributed, with very low proportions. However, the proportion of invalid answers is significantly higher than that of other error types. This suggests that the models generally understand the task structure correctly, but there are errors in specific geospatial data computations or parameter filling.

**Table 12. Error Type Distribution in JavaScript Code Generation Across Models**

| Model | Syntax Error | Attribute or Parameter Error | OutputType Error | Invalid Error | Runtime Error | Other Error |
|---|---|---|---|---|---|---|
| **General Non-Reasoning** | | | | | | |
| Deepseek-V3 | 7.20 | 17.42 | 2.62 | 72.04 | 0.00 | 0.71 |
| GPT-4.1 | 8.78 | 14.48 | 1.54 | 74.76 | 0.00 | 0.45 |
| GPT-4.1-mini | 12.51 | 16.93 | 1.75 | 68.81 | 0.00 | 0.00 |
| Qwen2.5-3B | 20.53 | 6.50 | 14.68 | 58.13 | 0.00 | 0.15 |
| Qwen2.5-7B | 32.46 | 8.58 | 12.42 | 46.22 | 0.09 | 0.23 |
| Qwen2.5-32B | 6.95 | 9.47 | 15.91 | 65.79 | 0.20 | 1.69 |
| Qwen3-4B | 14.01 | 8.94 | 21.94 | 54.81 | 0.07 | 0.24 |
| Qwen3-8B | 9.26 | 9.47 | 21.57 | 59.54 | 0.13 | 0.04 |
| Qwen3-32B | 7.38 | 12.80 | 9.88 | 69.68 | 0.00 | 0.27 |
| **General Reasoning** | | | | | | |
| o4-mini | 4.42 | 13.27 | 3.29 | 78.34 | 0.00 | 0.68 |
| QwQ-32B | 17.62 | 7.29 | 15.33 | 59.45 | 0.13 | 0.18 |
| Deepseek-R1 | 12.25 | 11.76 | 4.04 | 70.48 | 0.00 | 1.47 |
| **General Code Generation** | | | | | | |
| Deepseek-Coder-V2-16B | 22.80 | 9.27 | 11.02 | 56.16 | 0.00 | 0.75 |
| Qwen2.5-Coder-3B | 19.57 | 13.14 | 16.00 | 50.50 | 0.07 | 0.72 |
| Qwen2.5-Coder-7B | 15.52 | 10.27 | 19.74 | 54.39 | 0.04 | 0.04 |
| Qwen2.5-Coder-32B | 3.43 | 8.77 | 11.81 | 74.33 | 0.15 | 1.52 |
| Code-Llama-7B | 24.41 | 9.37 | 15.92 | 50.25 | 0.00 | 0.06 |
| **Geospatial Code Generation** | | | | | | |
| GeoCode-GPT | 21.54 | 7.16 | 16.44 | 54.73 | 0.05 | 0.08 |

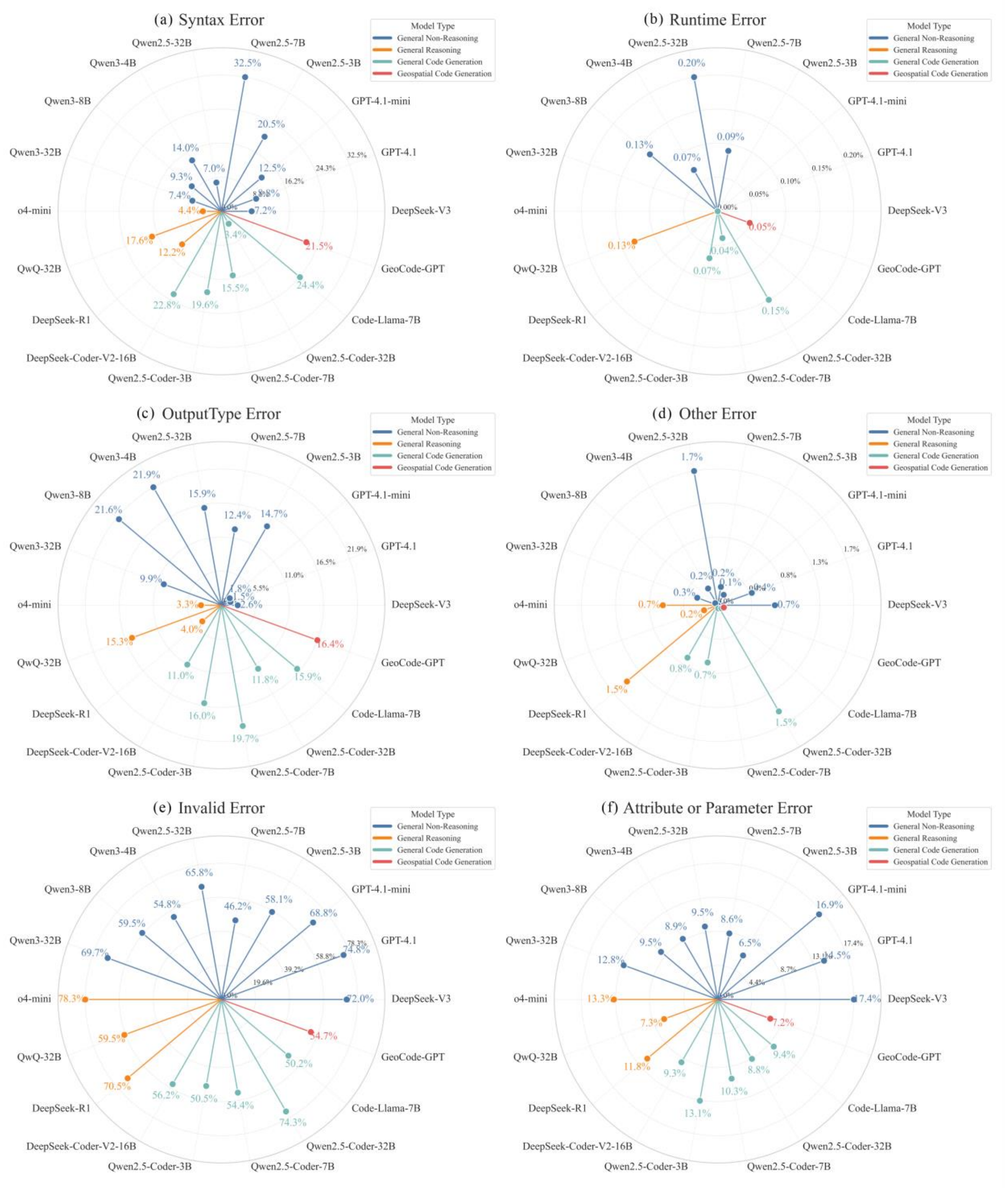

**Figure 14. Radar Chart of Different Error Types Observed in Model Outputs**

**Figure 14** shows the distribution of error types in function generation tasks across different models. Overall, Invalid Error is the most common error type, with over half of the models exhibiting more than 50% of this error, indicating that current models still have significant shortcomings in spatial semantic understanding and function call selection. This may be related to the large number of geospatial library functions, their similar semantics, and the lack of real geospatial code samples in the training data. General reasoning models have a relatively low proportion of Syntax Error and Output Type Error, demonstrating stronger capabilities in structural organization and type matching, likely benefiting from their logical chain construction and semantic reasoning abilities. GeoCode-GPT shows slight improvement in syntax and parameter errors compared to its base model, suggesting that domain fine-tuning can enhance the structural conformity of code to some extent. The overall proportion of Runtime Error and Other Error is small, indicating that most models have a certain level of competence in basic syntax correctness and executability.

### 6.5 Edge Case Testing

**Figure 15** shows the average boundary test pass rate for each model across five geospatial function libraries, measuring their generalization ability under extreme inputs. The results show that General Reasoning models performed the best, with an average pass rate of 71.0%, significantly outperforming General Non-Reasoning (60.8%), General Code Generation (58.1%), and GeoCode-GPT (41.4%). This indicates that models with stronger reasoning capabilities are better at handling complex or edge semantic conditions, making them more practical. There are also noticeable internal model differences. For example, Qwen2.5-3B has a pass rate of only 38.6%, significantly lower than similar models. In contrast, GPT-4.1-mini performs close to reasoning models (70.9%), demonstrating strong robustness. Model size has a positive impact on boundary capabilities, as both Qwen2.5 and Qwen2.5-Coder series show improved pass rates with increased parameter sizes. Overall, the GPT series shows stable boundary performance, with o4-mini ranking first with a 74.7% pass rate, further validating the effectiveness of reasoning structures in handling extreme tasks.

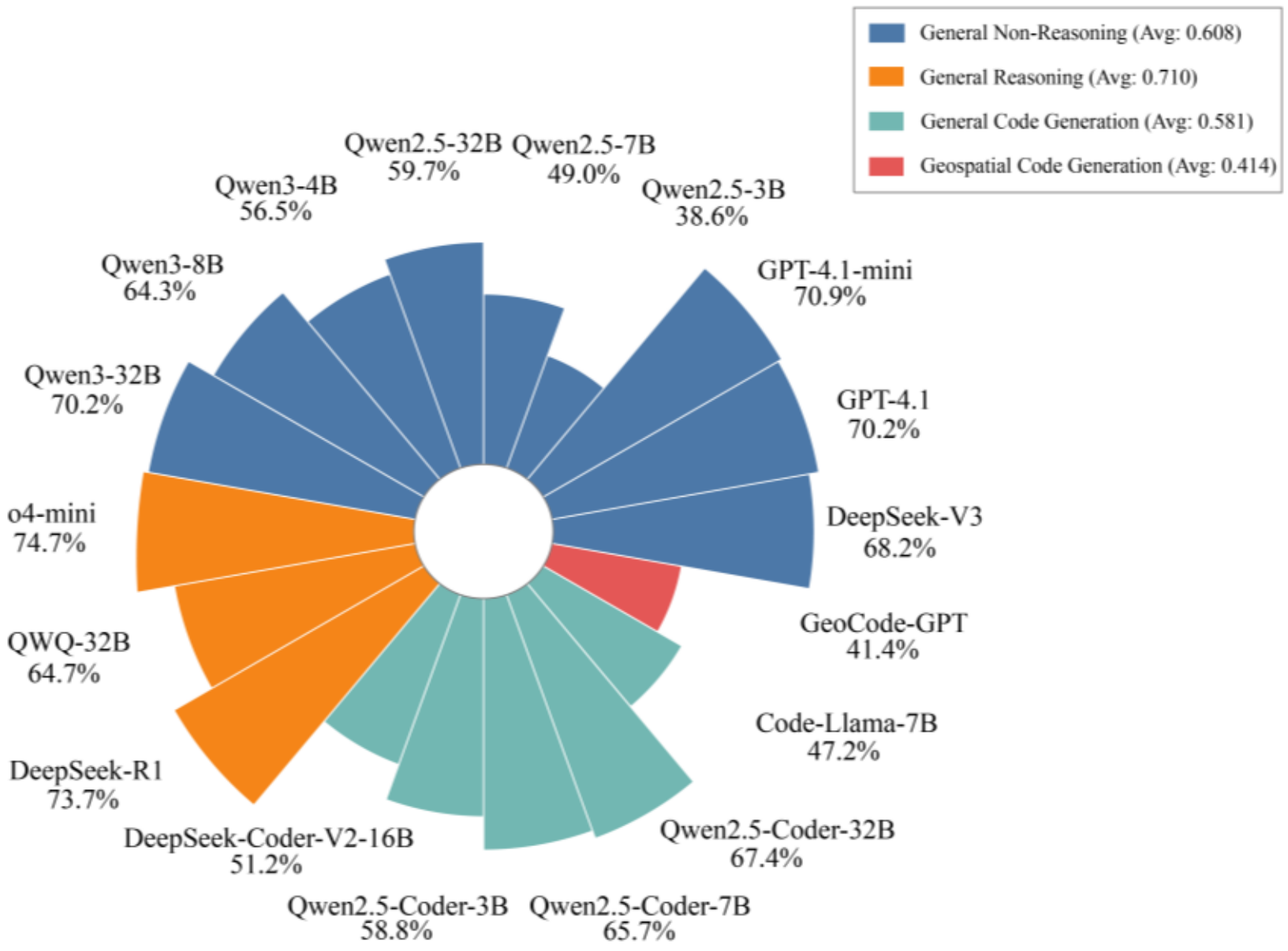

**Figure 15. Average Boundary Test Pass Rate per Model.**

### 6.6 Key Findings and Insights

This study systematically evaluates the comprehensive performance of 18 large language models in geospatial code generation tasks across four dimensions: accuracy, resource consumption, operational efficiency, and error types, based on the GeoJSEval evaluation framework. The main conclusions are as follows:

- Accuracy: The multi-round generation mechanism effectively alleviated the hallucination phenomenon in model generation, with Pass@3 significantly outperforming single-round generation. However, the improvement from Pass@3 to Pass@5 becomes more gradual, showing a clear diminishing marginal effect. Combining the CV and SA metrics, GPT-4.1, GPT-4.1-mini, and Qwen3-32B achieve a good balance between accuracy and stability.

- Resource consumption: General reasoning models consume significantly more tokens and have higher inference latencies compared to other models, showing clear efficiency shortcomings. In contrast, General Non-Reasoning models and Code Generation models have lower overall resource consumption and exhibit a higher computational cost-performance ratio.
- Operational efficiency: Qwen3-32B, Qwen2.5-Coder-32B, and Qwen2.5-Coder-7B show balanced performance across three dimensions: token usage, generation time, and output length, achieving the best efficiency. Although reasoning models have higher accuracy, their slower response times limit their application in scenarios that require real-time performance.
- Error type analysis: Invalid Answer is the most common error type, while the proportion of syntax and type errors is relatively low. This suggests that the model has basic syntax construction capabilities but still faces significant cognitive gaps in selecting spatial functions and matching semantics. Fine-tuning with real geospatial code data is needed to address these issues.
- Boundary test performance: General reasoning models exhibit the strongest boundary generalization ability, with average pass rates significantly higher than other categories. Increasing model size (e.g., Qwen2.5 series) significantly enhances boundary handling capabilities.
- Model differences: Among the DeepSeek series, DeepSeek-R1 achieves the highest accuracy but has the worst efficiency, making it suitable only for offline scenarios with extremely high quality requirements. DeepSeek-V3 shows excellent overall performance. DeepSeek-Coder-V2 ranks at the bottom across several metrics. The GPT series is generally stable, with o4-mini showing balanced performance across all three dimensions. GeoCode-GPT performs worse than its base model Code-Llama-7B, possibly due to its reliance on training data specific to the GEE platform, making it difficult to transfer to mainstream geospatial JS libraries.
- Model selection recommendations: For different task scenarios, model selection should be based on the Total_Rank metric, considering accuracy (P_Rank), stability (S_Rank), and efficiency (E_Rank) comprehensively. Models with high accuracy and efficiency (e.g., Qwen3-32B) are suitable for performance-sensitive tasks, models with high stability (e.g., GPT-4.1 series) are suitable for research or engineering scenarios, and cost-effective models supporting local deployment (e.g., Qwen2.5-Coder series) are more suitable for edge computing or large-scale applications.

## 7. Conclusion

This study introduces GeoJSEval, the first large language model automation evaluation framework specifically designed for JavaScript geospatial code generation tasks. It systematically examines the function understanding and code generation capabilities of mainstream LLMs across five core JS geospatial libraries (Turf.js, JSTS, Geolib, Leaflet, and OpenLayers), covering two key areas of capability: spatial analysis and map visualization. To this end, the standardized evaluation benchmark GeoJSEval-Bench was constructed, comprising 432 function-level test tasks and 2,071 structured sub-cases, covering 25 types of multimodal data structures, providing a comprehensive reflection of the diversity and complexity of geospatial function outputs. The evaluation tasks adopt a modular modeling strategy, relying on three core modules—function semantics, execution configuration, and evaluation parsing—to design test cases. Automated evaluation and comparison of model outputs are achieved through the answering and judging programs. The evaluation covers dimensions such as code accuracy, resource consumption, operational efficiency, and error types, encompassing 18 representative models across four categories: general language models, reasoning-augmented models, general code generation models, and geospatial-specific models. The evaluation results reveal the performance heterogeneity and capability boundaries of different models at the JavaScript geospatial library level. A reproducible, extensible, and structured geospatial code generation evaluation chain has been established, serving application scenarios in GIS, spatial visualization, and frontend development.

### 7.1 Significance and Advantages

The significance of this study is primarily reflected in three aspects. First, GeoJSEval is the first large language model evaluation framework focusing on JavaScript geospatial code generation tasks. It covers the complete process of task modeling, modular test design, benchmark construction, implementation of answering and judging programs, and evaluation metric setting, significantly reducing manual intervention and creating a structured, reproducible, and scalable automated evaluation system. Compared to traditional methods relying on manual screening and subjective judgment, GeoJSEval achieves a programmatic and standardized evaluation loop, improving the logical consistency and credibility of the evaluation results. Secondly, GeoJSEval conducts the first systematic evaluation of JavaScript geospatial code generation capabilities, filling the gap in previous research where the structure and invocation semantics of mainstream library functions like Turf.js and Leaflet were not modeled. Through empirical analysis, the study reveals the performance heterogeneity and capability shortcomings of mainstream models in tasks such as spatial computation and map rendering, providing clear directions and methodological support for future model fine-tuning, semantic enhancement, and system integration. Finally, GeoJSEval-Bench gathers 432 real API functions, 25 types of multimodal data structures, and 2,071 structured test cases, forming one of the most comprehensive resources for evaluating JavaScript geospatial code generation to date. This benchmark suite offers task diversity, structural coverage, and high reusability, making it widely applicable for geospatial agent development, multi-platform code evaluation, and toolchain construction, with significant potential for further dissemination and academic value.

### 7.2 Limitations and Future Work

Although this study has developed the first function-level automated evaluation framework for JavaScript geospatial code, GeoJSEval offers significant advantages in structural design and evaluation mechanisms, it still has several limitations that can be further expanded and improved in the future. First, in terms of evaluation coverage, although the current framework selects five representative mainstream libraries, it has not yet covered some specialized or poorly documented spatial computing and visualization libraries in the JavaScript ecosystem. Some functions were excluded due to complex encapsulation, lack of examples, or unclear semantics. In particular, APIs related to frontend interaction logic but not directly associated with map rendering have not been included in the evaluation, affecting the comprehensiveness of the assessment. In the future, it should be expanded to include community plugins, higher-level modules, and complex components to enhance adaptability to the JS geospatial function ecosystem and deepen the evaluation depth. Secondly, in terms of task modeling granularity, the current evaluation focuses on single function calls and does not cover typical logic in real-world development scenarios, such as function combinations, chained calls, and cross-module collaborations. In the future, a composite task set should be developed with a scenario-specific evaluation mechanism to better align with JavaScript geospatial analysis solving processes. Furthermore, there is room for further expansion in evaluation dimensions. For visual objects (e.g., Map and Layer), the evaluation primarily relies on structural field parsing and auxiliary function results. Although the automatic screenshot mechanism has been implemented, it has not yet been included in the scoring process due to rendering delays and the lack of image comparison tools, and it currently serves only as a reference for manual review. In the future, structure-aware image comparison algorithms can be introduced to incorporate visual behaviors such as rendering correctness and layer style overlay into the automated evaluation system.

## Author Contributions

Conceptualization, Huayi Wu and Shuyang Hou; methodology, Huayi Wu, Shuyang Hou and Ziqi Liu; software, Ziqi Liu and Guanyu Chen; validation, Guanyu Chen, Ziqi Liu and Haoyue Jiao; formal analysis, Guanyu Chen, Ziqi Liu and Haoyue Jiao; investigation, Guanyu Chen, Ziqi Liu; resources, Huayi Wu, Xuefeng Guan, Zhipeng Gui; data curation, Lutong Xie, Shaowen Wu, Guanyu Chen and Ziqi Liu ; writing—original draft preparation, Guanyu Chen and Shuyang Hou; writing—review and editing, Haoyue Jiao and Ziqi Liu; visualization, Haoyue Jiao, Guanyu Chen; supervision, Huayi Wu, Xuefeng Guan and Zhipeng Gui; project administration, Xuefeng Guan; funding acquisition, Zhipeng Gui
All authors have read and agreed to the published version of the manuscript.

## Funding

This research was funded by the National Natural Science Foundation of China, grant number 41971349. The APC was funded by the same source.

## Data Availability Statement

The experimental data used in this study can be downloaded from https://github.com/LiuZQ802/GeoJSEval.

## Conflicts of Interest

The authors declare no conflict of interest.

## Appendix A

**Figures A1** and **A2** present representative unit test cases from GeoJSEval-Bench, demonstrating direct evaluation of primitive output types and indirect evaluation of complex geospatial objects, respectively.

**3 Boolean**

**Function_header**

```
function G_isPointInLineBetweenTwoPoints(point, lineStart, lineEnd) {
    /**
     * The Geolib library has already been pre-loaded, so don't load it again.
     *
     * Use the given built-in functions from Geolib to complete the task as simply
as possible.
     *
     * Determines whether a given point lies exactly on the straight line segment
formed by two other points.
     *
     * @param {Object} point - The point to check in {latitude: number, longitude:
number} format
     * @param {Object} lineStart - Starting point of the line segment in {latitude:
number, longitude: number} format
     * @param {Object} lineEnd - Ending point of the line segment in {latitude:
number, longitude: number} format
     * @return {boolean} True if point lies on the line segment, false otherwise
     */
}
```

**Standard_code**

```
function G_isPointInLineBetweenTwoPoints(point, lineStart, lineEnd) {
    /**
     * The Geolib library has already been pre-loaded, so don't load it again.
     *
     * Use the given built-in functions from Geolib to complete the task as simply
as possible.
     *
     * Determines whether a given point lies exactly on the straight line segment
formed by two other points.
     *
     * @param {Object} point - The point to check in {latitude: number, longitude:
number} format
     * @param {Object} lineStart - Starting point of the line segment in {latitude:
number, longitude: number} format
     * @param {Object} lineEnd - Ending point of the line segment in {latitude:
number, longitude: number} format
     * @return {boolean} True if point lies on the line segment, false otherwise
     */
    return geolib.isPointInLine(point, lineStart, lineEnd);
}
```

**Parameter_list**

```
- params:
      point: !js |
        function get_Value() {
          const pointOnHorizontal = { latitude: 0, longitude: 5 };
          return pointOnHorizontal;
        }
      lineStart: !js |
        function get_Value() {
          const horizontalStart = { latitude: 0, longitude: 0 };
          return horizontalStart;
        }
      lineEnd: !js |
        function get_Value() {
          const horizontalEnd = { latitude: 0, longitude: 10 };
          return horizontalEnd;
        }
```

| | |
|---|---|
| **Output_type** | boolean |
| **Eval_methods** | [] |
| **Output_path** | G_isPointInLineBetweenTwoPoints _testcase.txt |

**Expected_answer**

```
true
```

**4 Feature**

**Function_header**

```
function T_feature(geometry, properties) {
    /**
     * The Turf library has already been pre-loaded, so don't load it again.
     *
     * Use the given built-in functions from Turf to complete the task as simply as
possible.
     *
     * Wraps the given GeoJSON Geometry(Object) in a GeoJSON Feature.
     * @param {Object} geometry - The GeoJSON Geometry object to be wrapped (e.g.,
Point, LineString, Polygon).
     * @param {Array<number>} [options.bbox] - Bounding box of the Feature [minX,
minY, maxX, maxY].
     * @return {GeoJSON.Feature} - The generated GeoJSON Feature containing the
input geometry.
     */
}
```

**Standard_code**

```
function T_feature(geometry, properties) {
    /**
     * The Turf library has already been pre-loaded, so don't load it again.
     *
     * Use the given built-in functions from Turf to complete the task as simply as
possible.
     *
     * Wraps the given GeoJSON Geometry(Object) in a GeoJSON Feature.
     * @param {Object} geometry - The GeoJSON Geometry object to be wrapped (e.g.,
Point, LineString, Polygon).
     * @param {Array<number>} [options.bbox] - Bounding box of the Feature [minX,
minY, maxX, maxY].
     * @return {GeoJSON.Feature} - The generated GeoJSON Feature containing the
input geometry.
     */
    return turf.feature(geometry, properties);
}
```

**Parameter_list**

```
- params:
      geometry:
        type: "LineString"
        coordinates: [[0, 0], [1, 1]]
      properties:
        name: "test line"
```

| | |
|---|---|
| **Output_type** | feature |
| **Eval_methods** | [] |
| **Output_path** | T_feature _testcase.geojson |

**Expected_answer**

```
{
  "type": "Feature",
  "geometry": {"type": "LineString", "coordinates": [[0, 0],[1, 1]]},
  "properties": {"name": "test line"}
}
```

**Figure A1.** Unit Test Example for Direct Evaluation of Primitive Output Types

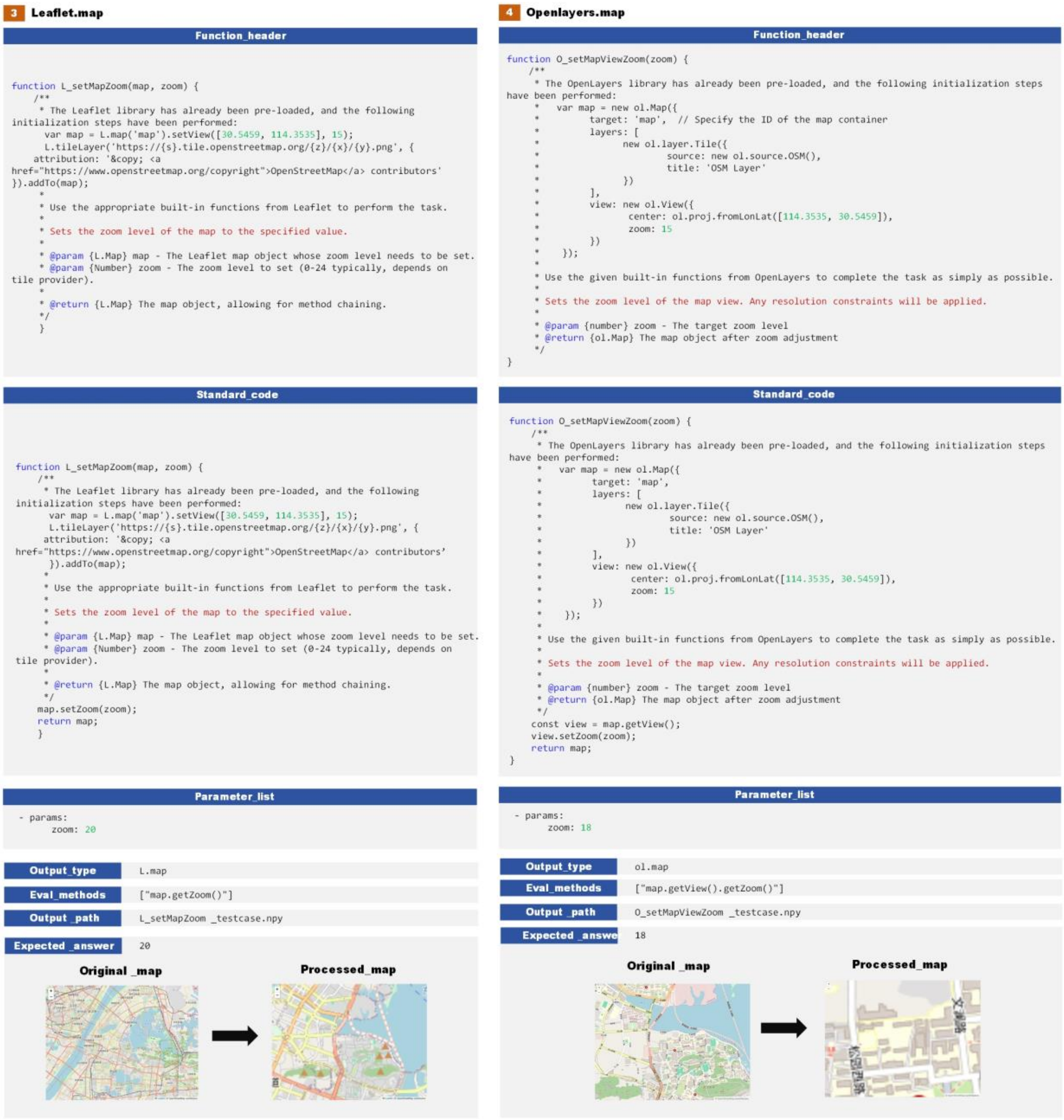

**Figure A2.** Unit Test Example for Indirect Evaluation of Complex Geospatial Objects

## References

Benveniste, A., Caillaud, B., & Le Guernic, P. (2000). Compositionality in dataflow synchronous languages: Specification and distributed code generation. *Information and Computation*, *163*(1), 125-171.

Brady, E. (2013). Idris, a general-purpose dependently typed programming language: Design and implementation. *Journal of functional programming*, *23*(5), 552-593.

Breunig, M., Bradley, P. E., Jahn, M., Kuper, P., Mazroob, N., Rösch, N., Al-Doori, M., Stefanakis, E., & Jadidi, M. (2020). Geospatial data management research: Progress and future directions. *ISPRS International Journal of Geo-Information*, *9*(2), 95.

Burnett, M. M., Baker, M. J., Bohus, C., Carlson, P., Yang, S., & Van Zee, P. (1995). Scaling up visual programming languages. *Computer*, *28*(3), 45-54.

Dai, W., Covvey, D., Alencar, P., & Cowan, D. (2009). Lightweight query-based analysis of workflow process dependencies. *Journal of Systems and Software*, *82*(6), 915-931.
Deng, Z., Ma, W., Han, Q.-L., Zhou, W., Zhu, X., Wen, S., & Xiang, Y. (2025). Exploring DeepSeek: A Survey on Advances, Applications, Challenges and Future Directions. *IEEE/CAA Journal of Automatica Sinica*, *12*(5), 872-893.
Dietrich, J., Jezek, K., & Brada, P. (2016). What Java developers know about compatibility, and why this matters. *Empirical Software Engineering*, *21*(3), 1371-1396.
Ding, Y., & Fotheringham, A. S. (1992). The integration of spatial analysis and GIS. *Computers, environment and urban systems*, *16*(1), 3-19.
Evangelidis, K., Papadopoulos, T., Papatheodorou, K., Mastorokostas, P., & Hilas, C. (2018). 3D geospatial visualizations: Animation and motion effects on spatial objects. *Computers & Geosciences*, *111*, 200-212.
Fang, Y., Huang, C., Su, Y., & Qiu, Y. (2020). Detecting malicious JavaScript code based on semantic analysis. *Computers & Security*, *93*, 101764.
Foerster, T., Stoter, J., & van Oosterom, P. (2012). On-demand base maps on the web generalized according to user profiles. *International Journal of Geographical Information Science*, *26*(1), 99-121.
Graser, A., & Olaya, V. (2015). Processing: A python framework for the seamless integration of geoprocessing tools in QGIS. *ISPRS International Journal of Geo-Information*, *4*(4), 2219-2245.
Gui, Z., Yang, C., Xia, J., Li, J., Rezgui, A., Sun, M., Xu, Y., & Fay, D. (2013). A visualization-enhanced graphical user interface for geospatial resource discovery. *Annals of GIS*, *19*(2), 109-121.
Gwak, J., Jung, J., Oh, R., Park, M., Rakhimov, M. A. K., & Ahn, J. (2019). A review of intelligent self-driving vehicle software research. *KSII Transactions on Internet and Information Systems (TIIS)*, *13*(11), 5299-5320.
Hochmair, H. H., Juhász, L., & Li, H. (2025). Advancing AI-Driven Geospatial Analysis and Data Generation: Methods, Applications and Future Directions. In (Vol. 14, pp. 56): MDPI.
Hou, D., Chen, J., & Wu, H. (2016). Discovering land cover web map services from the deep web with javascript invocation rules. *ISPRS International Journal of Geo-Information*, *5*(7), 105.
Hou, S., Jiao, H., Shen, Z., Liang, J., Zhao, A., Zhang, X., Wang, J., & Wu, H. (2025). Chain-of-programming (CoP): empowering large language models for geospatial code generation task. *International Journal of Dig*ital Earth, 18(1), 2509812.
Hou, S., Liang, J., Zhao, A., & Wu, H. GEE-OPs: an operator knowledge base for geospatial code generation on the Google Earth Engine platform powered by large language models. Geo-spatial Information Science, 1-22. https://doi.org/10.1080/10095020.2025.2505556
Hou, S., Shen, Z., Wu, H., Jiao, H., Liu, Z., Xie, L., Liu, C., Liang, J., Qing, Y., & Zhang, X. (2025). AutoGEEval++: A Multi-Level and Multi-Geospatial-Modality Automated Evaluation Framework for Large Language Models in Geospatial Code Generation on Google Earth Engine. arXiv preprint arXiv:2506.10365.
Hou, S., Shen, Z., Zhao, A., Liang, J., Gui, Z., Guan, X., Li, R., & Wu, H. (2025). GeoCode-GPT: A large language model for geospatial code generation. International Journal of Applied Earth Observation and Geoinformation, 104456.
Hou, S., Zhao, A., Liang, J., Shen, Z., & Wu, H. (2025). Geo-FuB: A method for constructing an Operator-Function knowledge base for geospatial code generation with large language models. Knowledge-Based Systems, 319, 113624. https://doi.org/https://doi.org/10.1016/j.knosys.2025.113624
Hu, T., Liu, T., Jarugumalli, V. S. D., Cheng, S., & Deng, C. (2025). FAIR principles in workflows: A GIScience workflow management system for reproducible and replicable studies. International *Journal of*

*Applied Earth Observation and Geoinformation*, *138*, 104477.
Jia, Y., Zhao, H., Niu, C., Jiang, Y., Gan, H., Xing, Z., Zhao, X., & Zhao, Z. (2009). A WebGIS-based system for rainfall-runoff prediction and real-time water resources assessment for Beijing. *Computers & Geosciences*, *35*(7), 1517-1528.
Le-Cong, T., Luong, D.-M., Le, X. B. D., Lo, D., Tran, N.-H., Quang-Huy, B., & Huynh, Q.-T. (2023). Invalidator: Automated patch correctness assessment via semantic and syntactic reasoning. *IEEE Transactions on Software Engineering*, *49*(6), 3411-3429.
Li, S., Dragicevic, S., Castro, F. A., Sester, M., Winter, S., Coltekin, A., Pettit, C., Jiang, B., Haworth, J., & Stein, A. (2016). Geospatial big data handling theory and methods: A review and research challenges. *ISPRS journal of Photogrammetry and Remote Sensing*, *115*, 119-133.
Li, Z., & Ning, H. (2023). Autonomous GIS: the next-generation AI-powered GIS. *International Journal of Digital Earth*, *16*(2), 4668-4686.
Liu, D. T., & Xu, X. W. (2001). A review of web-based product data management systems. *Computers in industry*, *44*(3), 251-262.
Marinoni, O. (2004). Implementation of the analytical hierarchy process with VBA in ArcGIS. *Computers & Geosciences*, *30*(6), 637-646.
Pakdil, M. E., & Çelik, R. N. (2022). Serverless geospatial data processing workflow system design. *ISPRS International Journal of Geo-Information*, *11*(1), 20.
Palomino, J., Muellerklein, O. C., & Kelly, M. (2017). A review of the emergent ecosystem of collaborative geospatial tools for addressing environmental challenges. *Computers, environment and urban systems*, *65*, 79-92.
Qu, Y., Huang, S., & Yao, Y. (2024). A survey on robustness attacks for deep code models. *Automated Software Engineering*, *31*(2), 65.
Quarati, A., De Martino, M., & Rosim, S. (2021). Geospatial open data usage and metadata quality. *ISPRS International Journal of Geo-Information*, *10*(1), 30.
Storey, M.-A. (2006). Theories, tools and research methods in program comprehension: past, present and future. *Software Quality Journal*, *14*(3), 187-208.
Sun, K., Zhu, Y., Pan, P., Hou, Z., Wang, D., Li, W., & Song, J. (2019). Geospatial data ontology: the semantic foundation of geospatial data integration and sharing. *Big Earth Data*, *3*(3), 269-296.
Wan, Y., Bi, Z., He, Y., Zhang, J., Zhang, H., Sui, Y., Xu, G., Jin, H., & Yu, P. (2024). Deep learning for code intelligence: Survey, benchmark and toolkit. *ACM Computing Surveys*, *56*(12), 1-41.
Wei, Q., Yao, Z., Cui, Y., Wei, B., Jin, Z., & Xu, X. (2024). Evaluation of ChatGPT-generated medical responses: a systematic review and meta-analysis. *Journal of biomedical informatics*, *151*, 104620.
Wu, H., Shen, Z., Hou, S., Liang, J., Jiao, H., Qing, Y., Zhang, X., Li, X., Gui, Z., & Guan, X. (2025). AutoGEEval: A Multimodal and Automated Evaluation Framework for Geospatial Code Generation on GEE with Large Language Models. *ISPRS International Journal of Geo-Information*.
Xu, F. F., Vasilescu, B., & Neubig, G. (2022). In-ide code generation from natural language: Promise and challenges. *ACM Transactions on Software Engineering and Methodology (TOSEM)*, *31*(2), 1-47.
Zhao, Q., Yu, L., Li, X., Peng, D., Zhang, Y., & Gong, P. (2021). Progress and trends in the application of Google Earth and Google Earth Engine. *Remote Sensing*, *13*(18), 3778.